\documentclass[preprint2]{aastex}

\usepackage{graphicx}
\usepackage{graphics}
\usepackage{amsmath}
\usepackage{txfonts}
\usepackage{epsfig}
\usepackage{natbib}
\usepackage{times}
\usepackage{amssymb}
\usepackage{color}

\newcommand\igrone{\object{IGR~J18457$+$0244}}
\newcommand\igrtwo{\object{IGR~J18462$-$0223}}
\newcommand\igrthree{\object{IGR~J18482$+$0049}}
\newcommand\igrfour{\object{IGR~J18532$+$0416}}
\newcommand\igrfive{\object{IGR~J18538$-$0102}}
\newcommand\ergcms{erg\,cm$^{-2}$\,s$^{-1}$}

\newcommand\ergs{erg\,s$^{-1}$}
\newcommand\cmsq{cm$^{-2}$}
\newcommand\integ{{\it{INTEGRAL}}}

\newcommand\swift{{\it{Swift}}}

\newcommand\xmm{{\it{XMM-Newton}}}

\newcommand\nh{$N_\mathrm{H}$}

\shorttitle{\emph{XMM-Newton} observations of five IGRs towards the Scutum Arm}
\shortauthors{Bodaghee et al.}

\begin{document}

\title{\emph{XMM-Newton} observations of five \emph{INTEGRAL} sources\\located towards the Scutum Arm}

\author{A. Bodaghee and J. A. Tomsick}
\affil{Space Sciences Laboratory, 7 Gauss Way, University of California, Berkeley, CA 94720, USA}
\email{bodaghee@ssl.berkeley.edu}

\and

\author{J. Rodriguez}
\affil{Laboratoire AIM, CEA/IRFU - Universit\'e Paris Diderot - CNRS/INSU, \\ CEA DSM/IRFU/SAp, Centre de Saclay, F-91191 Gif-sur-Yvette, France}

\begin{abstract}
Results are presented for \xmm\ observations of five hard X-ray sources discovered by \integ\ in the direction of the Scutum Arm. Each source received $\gtrsim$20\,ks of effective exposure time. We provide refined X-ray positions for all 5 targets enabling us to pinpoint the most likely counterpart in optical/infrared archives. Spectral and timing information (much of which are provided for the first time) allow us to give a firm classification for \object{IGR~J18462$-$0223} and to offer tentative classifications for the others. For \object{IGR~J18462$-$0223}, we discovered a coherent pulsation period of 997$\pm$1\,s which we attribute to the spin of a neutron star in a highly-obscured (\nh\ $= 2\times10^{23}$\,\cmsq) high-mass X-ray binary (HMXB). This makes \object{IGR~J18462$-$0223} the seventh supergiant fast X-ray transient (SFXT) candidate with a confirmed pulsation period. \object{IGR~J18457$+$0244} is a highly-absorbed (\nh\ $= 8\times10^{23}$\,\cmsq) source in which the possible detection of an iron line suggests an active galactic nucleus (AGN) of type Sey-2 situated at $z = 0.07(1)$. A periodic signal at 4.4\,ks could be a quasi-periodic oscillation which would make \object{IGR~J18457$+$0244} one of a handful of AGN in which such features have been claimed, but a slowly-rotating neutron star in an HMXB can not be ruled out. \object{IGR~J18482$+$0049} represents a new obscured HMXB candidate with \nh\ $= 4\times10^{23}$\,\cmsq. We tentatively propose that \object{IGR~J18532$+$0416} is either an AGN or a pulsar in an HMXB system. The X-ray spectral properties of \igrfive\ are consistent with the AGN classification that has been proposed for this source.
\end{abstract}


%%__________________________________________________________________
\section{Introduction}

Surveys by \integ\ have enabled the discovery of hundreds of new high-energy sources \citep[e.g.,][]{bir10,kri10}. While \integ\ has proven adept at finding new sources, the position error radii are on the order of a few arcminutes. These are clearly too large to permit the identification of a single counterpart in the optical and infrared bands. Establishing the nature of the optical/IR counterpart is a crucial element in helping to categorize an \integ\ Gamma-Ray source (IGR)\footnote{a comprehensive list of IGRs and their properties can be found at \texttt{http://irfu.cea.fr/Sap/IGR-Sources}} into one of the many groups of high-energy emitters. 

%__________________________________________________________________GAL NH
\begin{figure*}[!t] 
\centering
\includegraphics[width=\textwidth,angle=0]{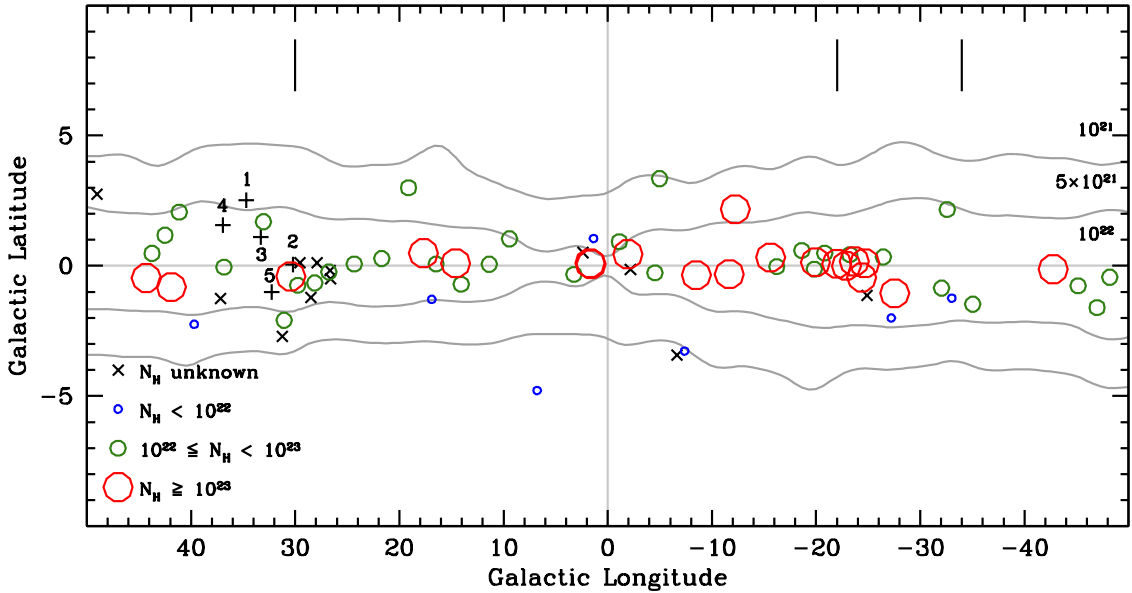}
\caption{The spatial distribution of HMXBs in Galactic coordinates with symbol size proportional to the X-ray measured column density as reported in the literature. The contours represent cumulative line-of-sight absorption levels of $10^{21}$, $5\times 10^{21}$, and $10^{22}$\,cm$^{-2}$ \citep{dic90}. The vertical bars denote the tangents to the (left to right) Scutum, Inner Perseus, and Norma arms from \citet{val08}. The five targets in this study (which are unclassified or tentatively classified) are labeled with crosses and are numbered 1 through 5 in order of increasing R.A. (see Table\,\ref{tab_log}). The \nh\ values are known for some of the targets (see Section\,\ref{sec_res}).}
\label{fig_gal_nh}
\end{figure*}

Hence, the classification of IGRs depends on subsequent observations with X-ray focusing telescopes which provide position accuracies of a few arcseconds \citep[e.g.,][]{rod10}. What these follow-up observations have shown is that some IGRs belong to a previously-rare subclass of high-mass X-ray binary (HMXB) called supergiant X-ray binaries \citep[SGXBs: e.g.,][and references therein]{wal06}. Of the 50 IGR HMXBs, 26 of them are SGXBs representing new additions to a subclass that once contained less than a dozen members \citep{Liu00}: the most prominent examples of which are \object{Vela~X-1} and GX\,{301$-$2}. A common characteristic of SGXBs is their high intrinsic column (\nh\ $\ge 5\times10^{22}$\,\cmsq) which suggests that the compact accretor is embedded in the dense winds shed by its supergiant companion star. The most extreme case is \object{IGR~J16318$-$4848} with the largest \nh\ of any X-ray source known in the Galaxy \citep[\nh\ $\sim 2\times10^{24}$\,\cmsq, ][]{mat03}. The typical \nh\ value of these systems is a few $\times10^{23}$\,\cmsq.

The spatial distribution of obscured HMXBs in the inner Milky Way appears to be asymmetric about the Galactic Center (GC) \citep{bod07}. In the inner quadrant of the Galactic Plane (Fig.\,\ref{fig_gal_nh}), there are 37 HMXBs (with and without \nh\ measurements) at positive longitudes, and an equivalent number (34) at negative longitudes. Among HMXBs whose \nh\ values are known to be less than $10^{23}$\,\cmsq, the left-right distribution is also symmetrical about the GC: 21 vs. 18. Thus, the asymmetry is only evident for the most obscured systems (\nh\ $\geq 10^{23}$\,\cmsq): there are around half as many obscured HMXBs (7) at positive longitudes as there are at negative longitudes (13). On the other hand, we have an incomplete picture of the obscured HMXB population in the direction of the Scutum Arm. There are unclassified IGRs (including HMXB candidates) in this region. If some of these are shown to be obscured HMXBs, then it will level the Galactic distribution, which might indicate that the asymmetry was the result of an observational bias rather than being due to, e.g., a possible evolutionary difference between the hard X-ray populations of the arms.

%__________________________________________________________________Journal
%
\begin{table*}[!t] 
\caption{Journal of \xmm\ observations. The effective exposure time represents the time dedicated to the source position (with EPIC-pn) after the filtering and cleaning processes.}
\vspace{2mm}
\begin{tabular}{ l c c c c c}
\hline
\hline
Target	      				& Spacecraft Rev. 	& 	Obs. ID 		& Start Time (UTC)		& End Time (UTC)		& Eff. Exp. (ks)		\\	
\hline
\object{IGR~J18457$+$0244}	& 2075			& 0651680201		& 2011-04-09T05:18:17	& 2011-04-09T12:41:58	& 25.036					\\	
\object{IGR~J18462$-$0223}	& 2080			& 0651680301		& 2011-04-18T08:52:18	& 2011-04-18T17:44:32	& 31.937					\\
\object{IGR~J18482$+$0049}	& 2079			& 0651680401		& 2011-04-16T18:39:11	& 2011-04-17T01:06:24	& 23.236					\\
\object{IGR~J18532$+$0416}	& 2077			& 0651680501		& 2011-04-12T12:10:49	& 2011-04-12T17:56:41	& 20.752					\\
\object{IGR~J18538$-$0102}	& 2081			& 0651680101		& 2011-04-20T14:44:35	& 2011-04-20T21:48:30	& 25.437					\\
\hline
\end{tabular}
\label{tab_log}
\end{table*}

In order to help complete, at least partially, the sample of obscured HMXBs, we obtained \xmm\ observations of five unclassified (or tentatively classified) IGRs that are located in the direction of the Scutum Arm. Their persistent emission above 20\,keV, lack of known X-ray counterpart below 10\,keV, and location close to the Galactic Plane make them good candidates for being obscured HMXBs. We present our observations and analysis methods in Section\,\ref{sec_obs}. Results for individual sources are discussed and summarized in Sections\,\ref{sec_res}--\ref{sec_sum}.

%__________________________________________________________________IMG:XMM
\begin{figure*}[!t] 
\centering
\includegraphics[width=\textwidth,angle=0]{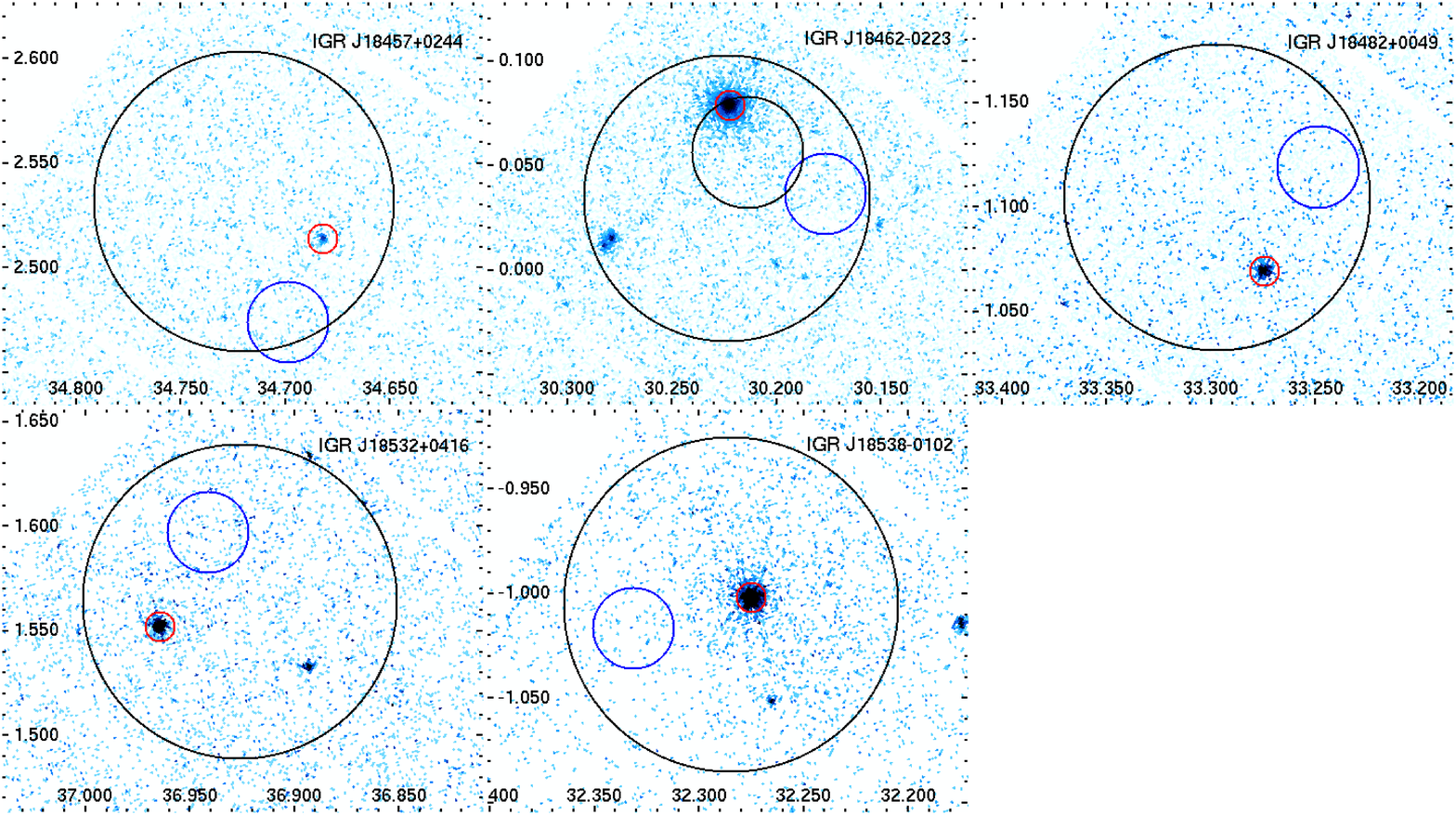}
\caption{EPIC-MOS1 images (0.5--10\,keV) in Galactic coordinates of the five targets in this study. The black circles represent the \emph{INTEGRAL}-ISGRI 90\%-confidence circles: the large circles are from \citet{bir10}, while for IGR~J18462$-$0223, the more accurate ISGRI position from \citet{gre10} is also shown. The source and background extraction regions are presented as red and blue circles, respectively. North is up and East is left. }
\label{fig_img}
\end{figure*}

%%__________________________________________________________________
\section{Observations \& Analysis}
\label{sec_obs}

Our target list consists of five hard X-ray sources located towards the Scutum Arm. These sources were observed by the \xmm\ \citep{jan01,str01,tur01} telescope during April, 2011, for approximately 30\,ks each (Observation IDs: 0651680101--0651680501; PI: Bodaghee). Table \ref{tab_log} provides the observation logs. All data reduction and analysis steps were performed using \texttt{HEASOFT} 6.11 and the Science Analysis System (SAS) 11.0.0.

%__________________________________________________________________Posns
%
\begin{table*}[!t] 
\caption{X-ray positions (J2000) from \xmm\ for the IGR sources in this program. The positions (90\% confidence radius of 2$\overset{\prime\prime}{.}$5) represent an average of the positions found by running \texttt{edetect} on pn, MOS1, and MOS2. The nearest infrared counterpart from 2MASS \citep{skr06} is listed, as is its offset with the X-ray position (in arcseconds). }
\vspace{2mm}
\begin{tabular}{ l c c c c l c }
\hline
\hline
Source Name	      			& R.A. 		& Dec.	& $l$ & $b$ & counterpart candidate & offset ($^{\prime\prime}$)	\\	
\hline
\object{IGR~J18457$+$0244}	& $18^{\mathrm{h}} 45^{\mathrm{m}} 40\overset{\mathrm{s}}{.}30$		& $+02^{\circ} 42^{\prime} 11\overset{\prime\prime}{.}2$	& 34.682 & $+$2.514 & \object{2MASS~J18454039$+$0242088} & 2.7 \\

\object{IGR~J18462$-$0223}	& $18^{\mathrm{h}} 46^{\mathrm{m}} 12\overset{\mathrm{s}}{.}68$		& $-02^{\circ} 22^{\prime} 29\overset{\prime\prime}{.}3$	& 30.223 & $+$0.079 & \object{2MASS~J18461279$-$0222261} & 3.4 \\		
\object{IGR~J18482$+$0049}	& $18^{\mathrm{h}} 48^{\mathrm{m}} 15\overset{\mathrm{s}}{.}32$		& $+00^{\circ} 47^{\prime} 34\overset{\prime\prime}{.}9$	& 33.275 & $+$1.070 & \object{2MASS~J18481540$+$0047332} & 2.0 \\		
\object{IGR~J18532$+$0416}	& $18^{\mathrm{h}} 53^{\mathrm{m}} 15\overset{\mathrm{s}}{.}83$		& $+04^{\circ} 17^{\prime} 48\overset{\prime\prime}{.}5$	& 36.965 & $+$1.553 & \object{2MASS~J18531602$+$0417481} & 2.9 \\		
\object{IGR~J18538$-$0102}	& $18^{\mathrm{h}} 53^{\mathrm{m}} 48\overset{\mathrm{s}}{.}42$		& $-01^{\circ} 02^{\prime}  28\overset{\prime\prime}{.}3$	& 32.275 & $-$1.002 & \object{2MASS~J18534847$-$0102295} & 1.6 \\		
\hline
\end{tabular}
\label{tab_posn}
\end{table*}

We reprocessed the MOS and pn events files using \texttt{emproc} and \texttt{epproc}, respectively. To identify epochs with a high particle background, we used \texttt{evselect} to create single-event (i.e., \texttt{PATTERN}$==$0) light curves above 10\,keV for MOS1/2 and between 10 and 12\,keV for pn which covered the full field of view (FOV) of each instrument. For the spectral analysis, we excluded epochs in which the particle background was higher than 1 count per second (cps). The good time intervals (GTIs) that remained were then used in \texttt{evselect} to produce filtered event files for each instrument. Time stamps from the satellite reference frame were converted to that of the Solar System's barycenter with the \texttt{barycen} tool. 

%__________________________________________________________________LC BG
\begin{figure*}[!t] 
\centering
\includegraphics[width=6cm,angle=0]{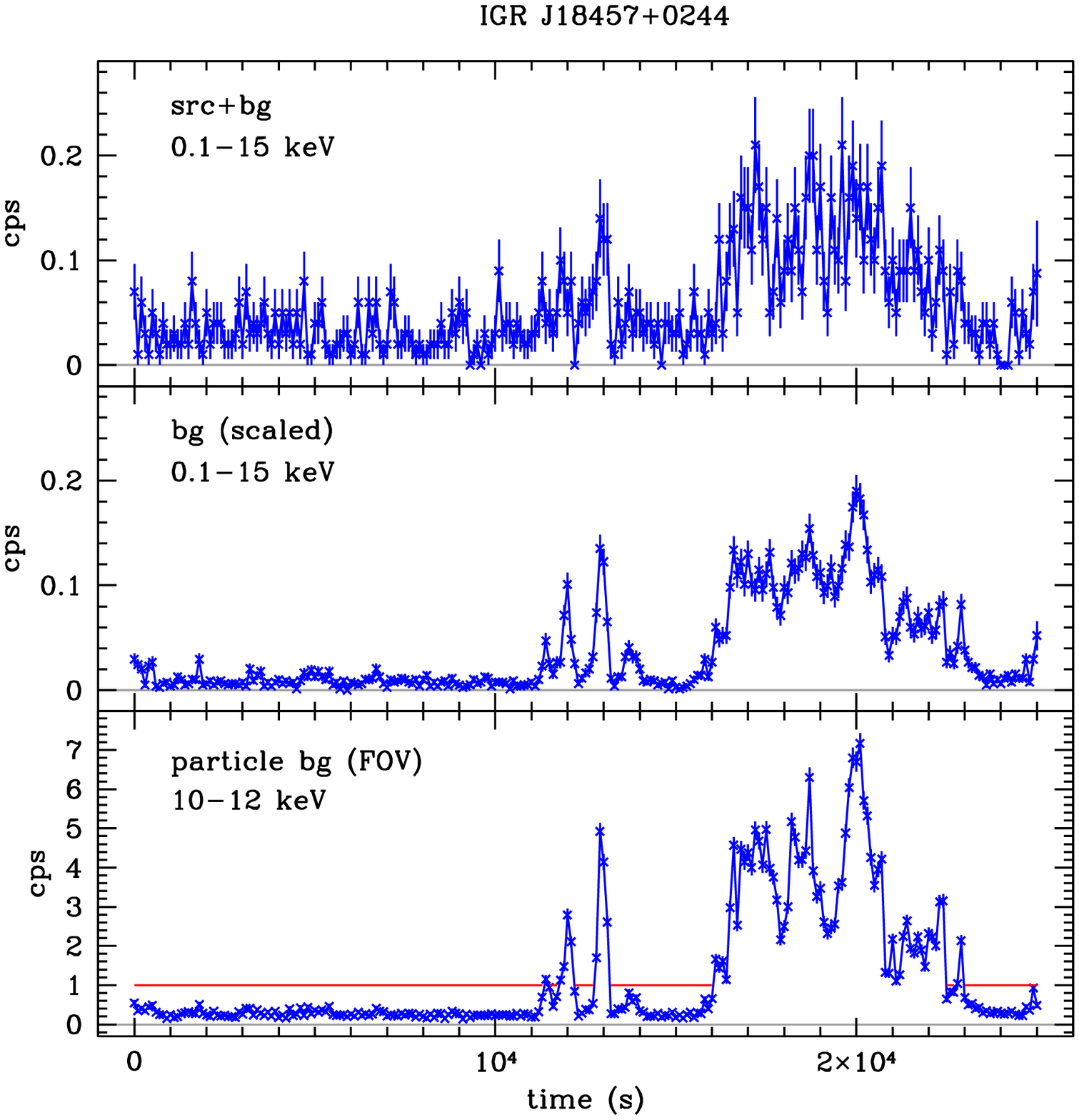}\includegraphics[width=6cm,angle=0]{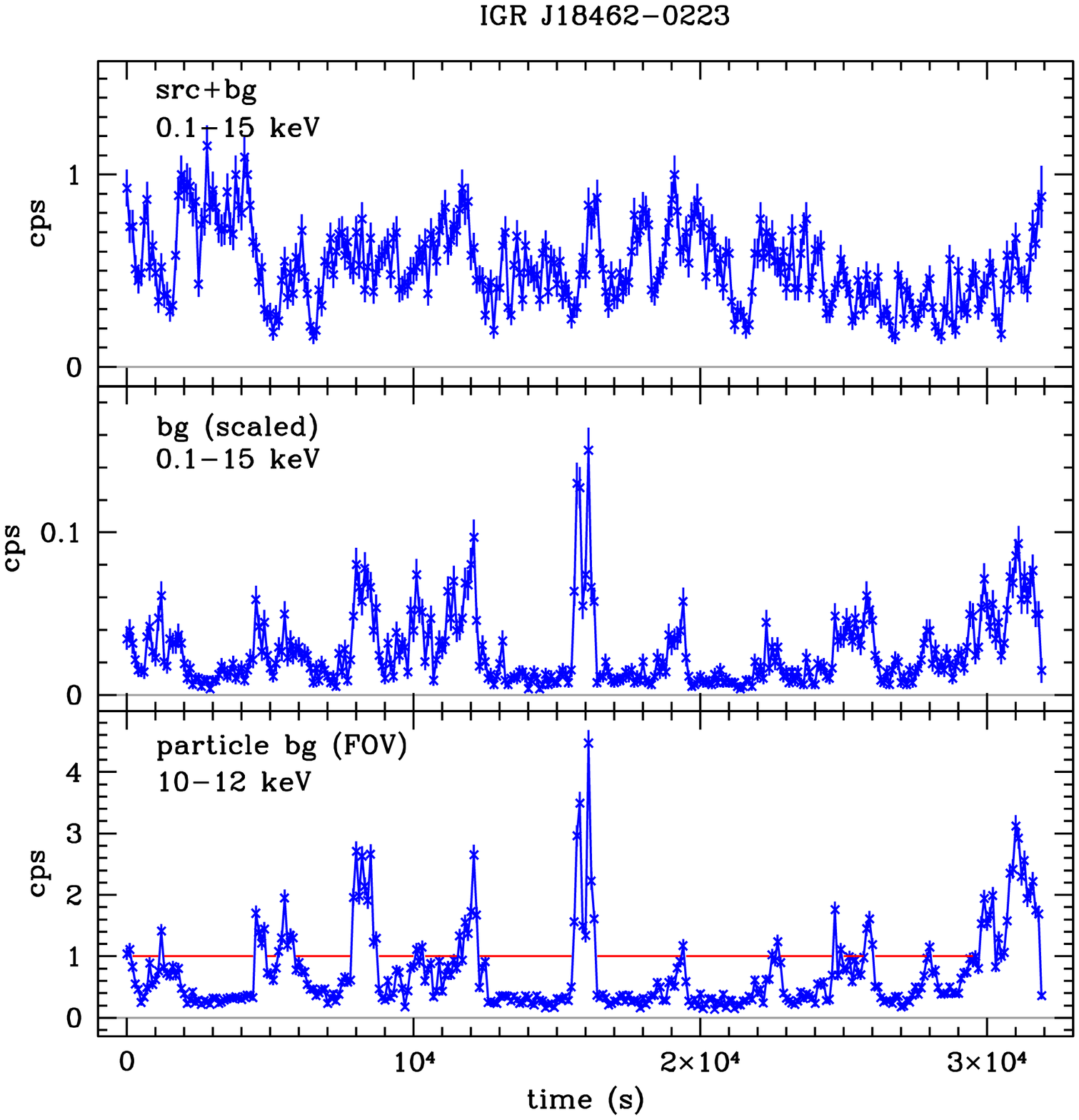}\includegraphics[width=6cm,angle=0]{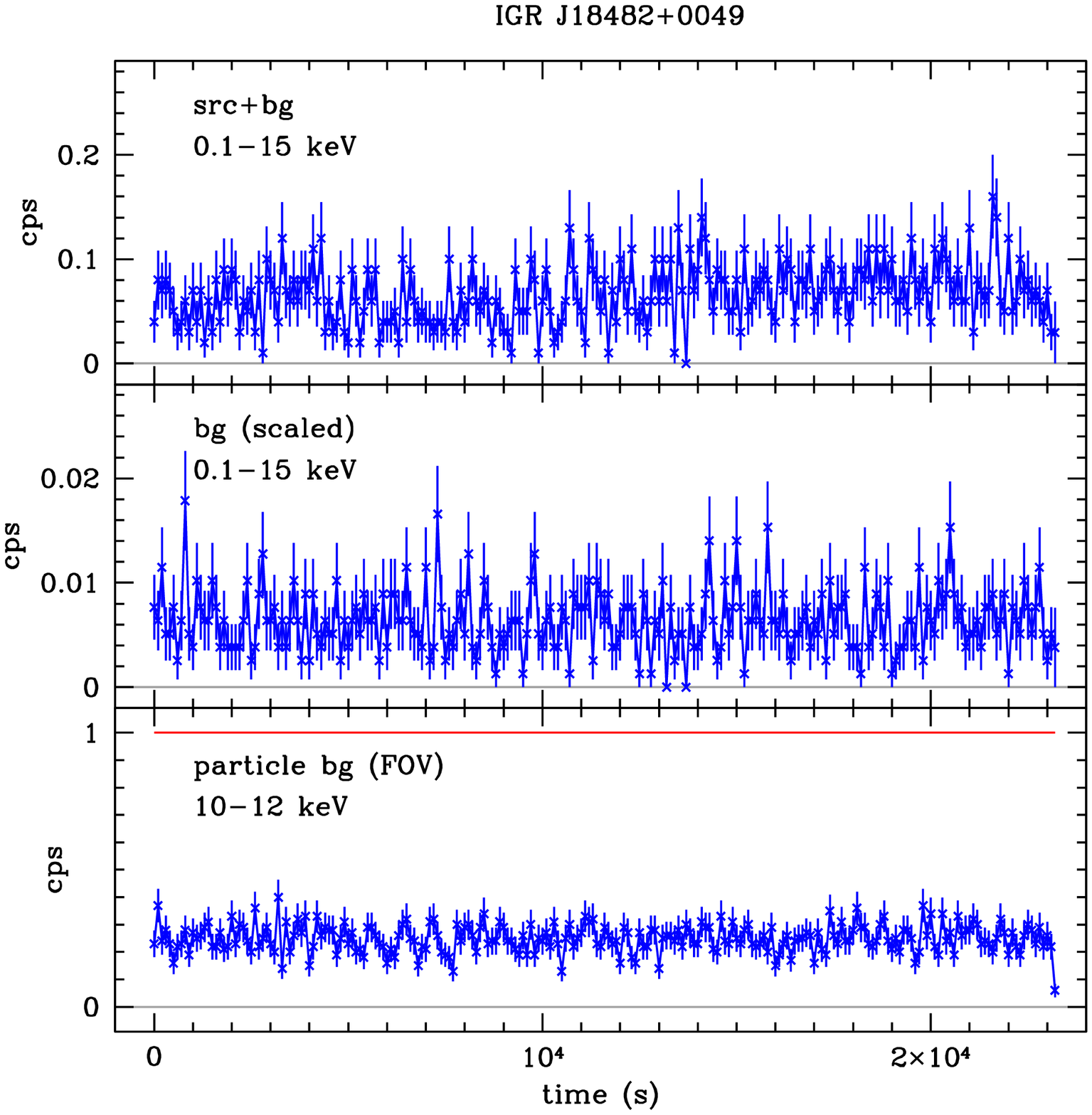}
\includegraphics[width=6cm,angle=0]{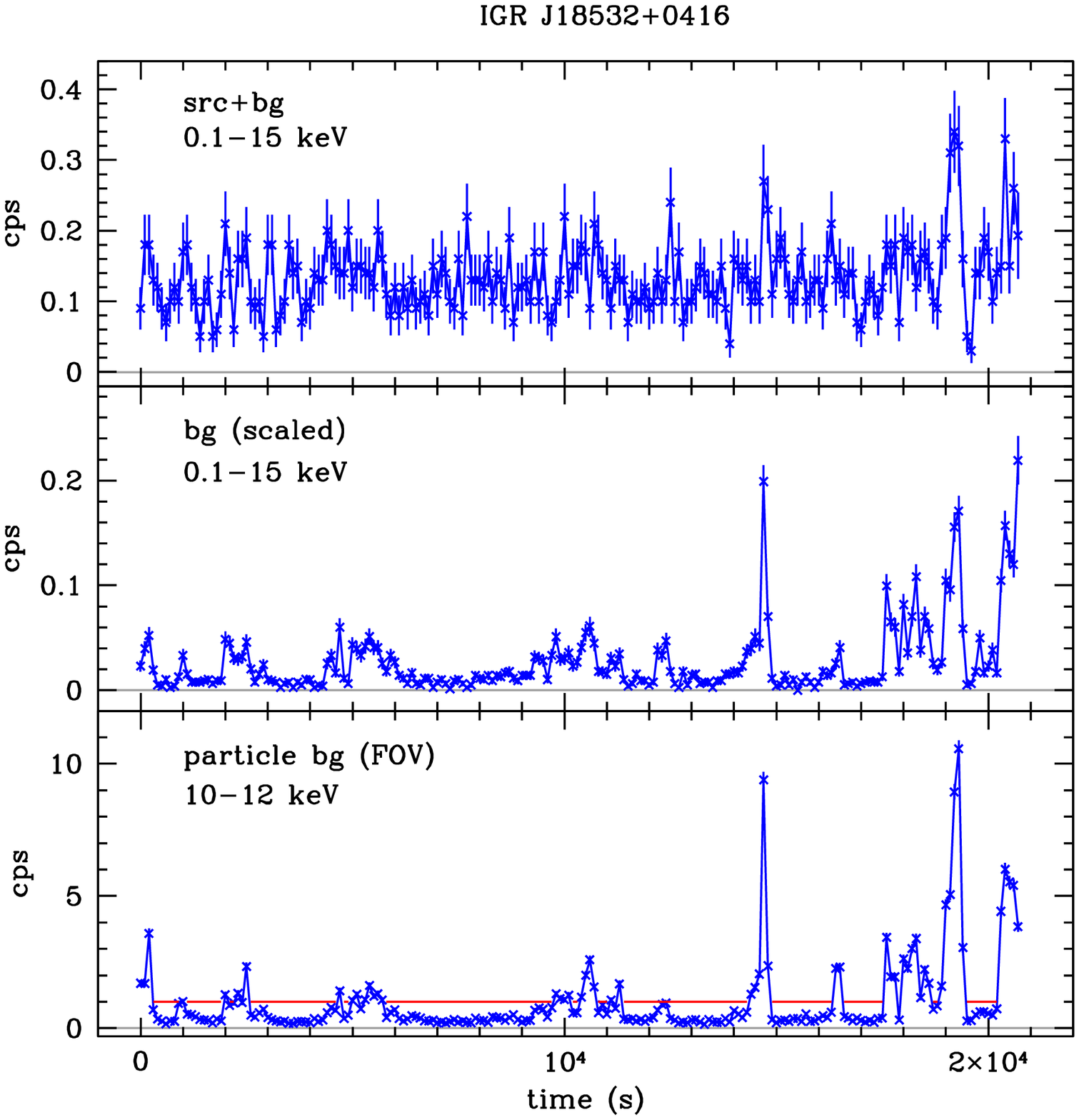}\includegraphics[width=6cm,angle=0]{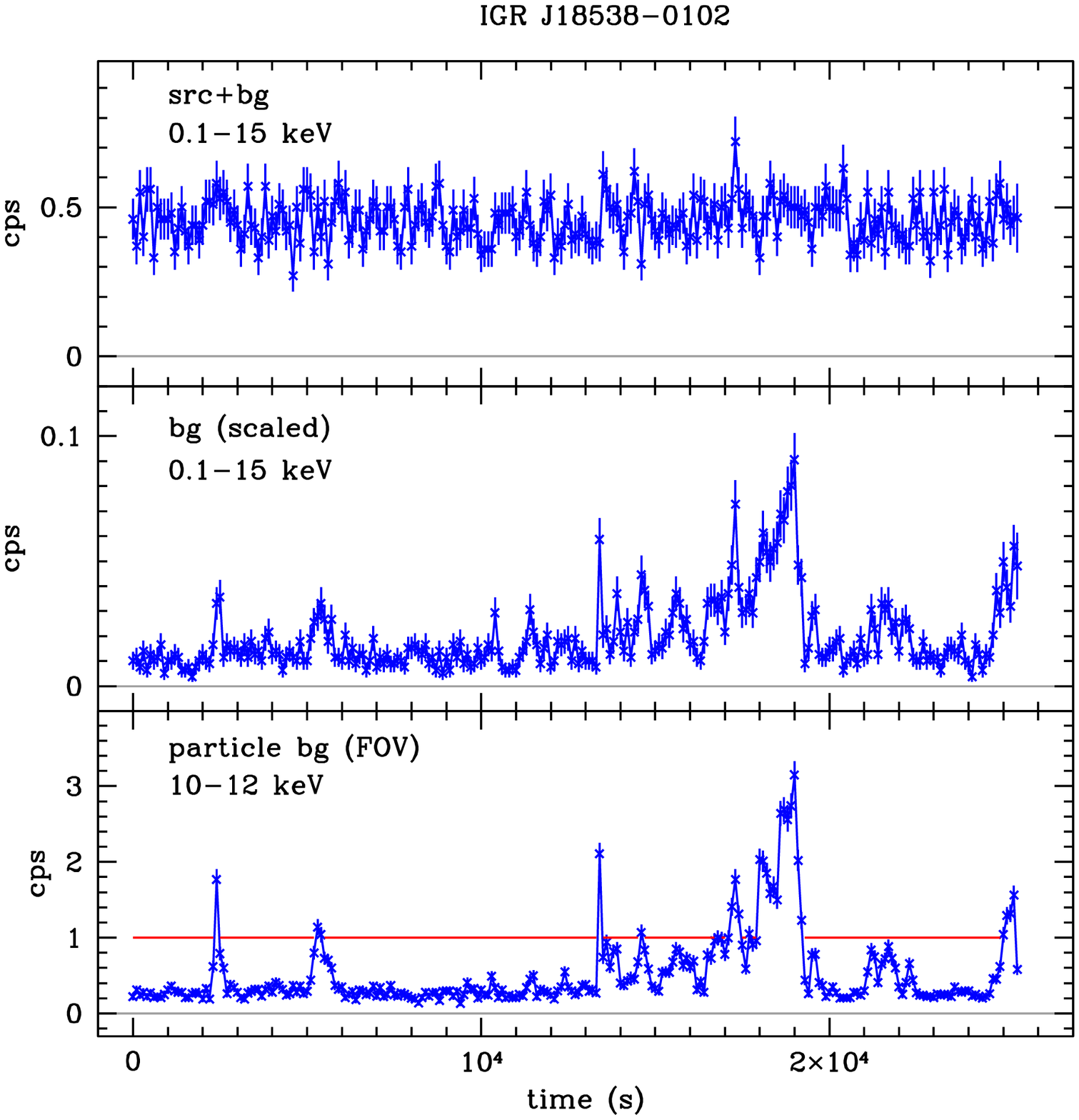}
\caption{EPIC-pn light curves with a time time resolution of 100\,s. In each figure, the top and middle panels present the raw count rates in the 0.13--15-keV range from the source (not corrected for the background) and background extraction regions (scaled to the size of the source region). The bottom panel shows the count rate between 10 and 12\,keV in the full field of view (FOV) attributed to the particle background. Only time intervals during which the particle background rate was below 1\,cps (represented by the red line) are considered in the spectral analysis.}
\label{lc_bg}
\end{figure*}

From these cleaned event lists, we generated images for MOS1/2 and for pn over the full FOV and covering the full spectrum of each instrument. The source extraction region consisted of a circle with a radius of 25$^{\prime\prime}$ ($=$ 500\,pixels in the unbinned image) centered on the brightest pixel. This radius was chosen to be wide enough to encompass most of the source photons (this corresponds to an encircled energy fraction of $\sim$80\% for the PSF from MOS and pn according to the SAS user's manual\footnote{\texttt{http://xmm.esac.esa.int/external/xmm\_user\_support/\linebreak documentation/uhb\_2.5/node17.html}}). However, since some of our targets are located near the edge of a CCD, we also wanted this radius to be restrictive enough so as to avoid collecting photons from an adjacent CCD. For the background extraction region, we used a circle of radius equal to 70$^{\prime\prime}$ (1400\,pixels) from a source-free region located on the same CCD as the target source. Whenever possible, this background region was chosen to be more than 1$^{\prime}$ away from the target source, at an equivalent off-axis angle as the source region in MOS, or centered at the same \texttt{RAW\_Y} pixel in pn (i.e., the center of the background region is shifted horizontally in the detector space with respect to the center of the source region). 

Figure\,\ref{fig_img} presents the MOS1 images of the fields centered on the \integ-ISGRI positions of the five targets from \citet{bir10}. We ran \texttt{edetect} on the cleaned event files of each instrument and produced lists of detected sources in several energy bands. The output list from \texttt{edetect} includes a dozen or more source detections, a few of which are located inside the 90\%-confidence ISGRI error circle. Spectra were extracted for all sources detected inside the most accurate ISGRI position available: the radius of the error circle is 1$\overset{\prime}{.}$6 for \igrtwo\ \citep{gre10}, and $\sim$4$^{\prime}$ for the others \citep{bir10}. Within each ISGRI circle, we identified a single \xmm\ counterpart (which happened to be the brightest one) whose spectrum was consistent with the source being an IGR, i.e., a hard power law ($\Gamma\sim $0--2) with some photoelectric absorption (\nh\ $\geq 10^{21}$\,\cmsq). We were also able to compare the spectral parameters (and coordinates) with those from previous soft X-ray observations (Section\,\ref{sec_res}).

Once the correct X-ray counterpart was identified for each IGR source, its coordinates from MOS1/2 and pn were averaged and this position is reported in Table\,\ref{tab_posn} (equinox 2000.0 is used throughout this work). According to the latest calibration documents (XMM-SOC-CAL-TN-0018\footnote{\texttt{http://xmm2.esac.esa.int/external/xmm\_sw\_cal/calib}}), the position uncertainty for EPIC is 1$\overset{\prime\prime}{.}$5 (at 68\% confidence) which is dominated by systematics. Thus, we have adopted a value of 2$\overset{\prime\prime}{.}$5 (at 90\% confidence) for the source position uncertainty.

\begin{figure*}[!t] 
\centering
\includegraphics[width=6cm,angle=0]{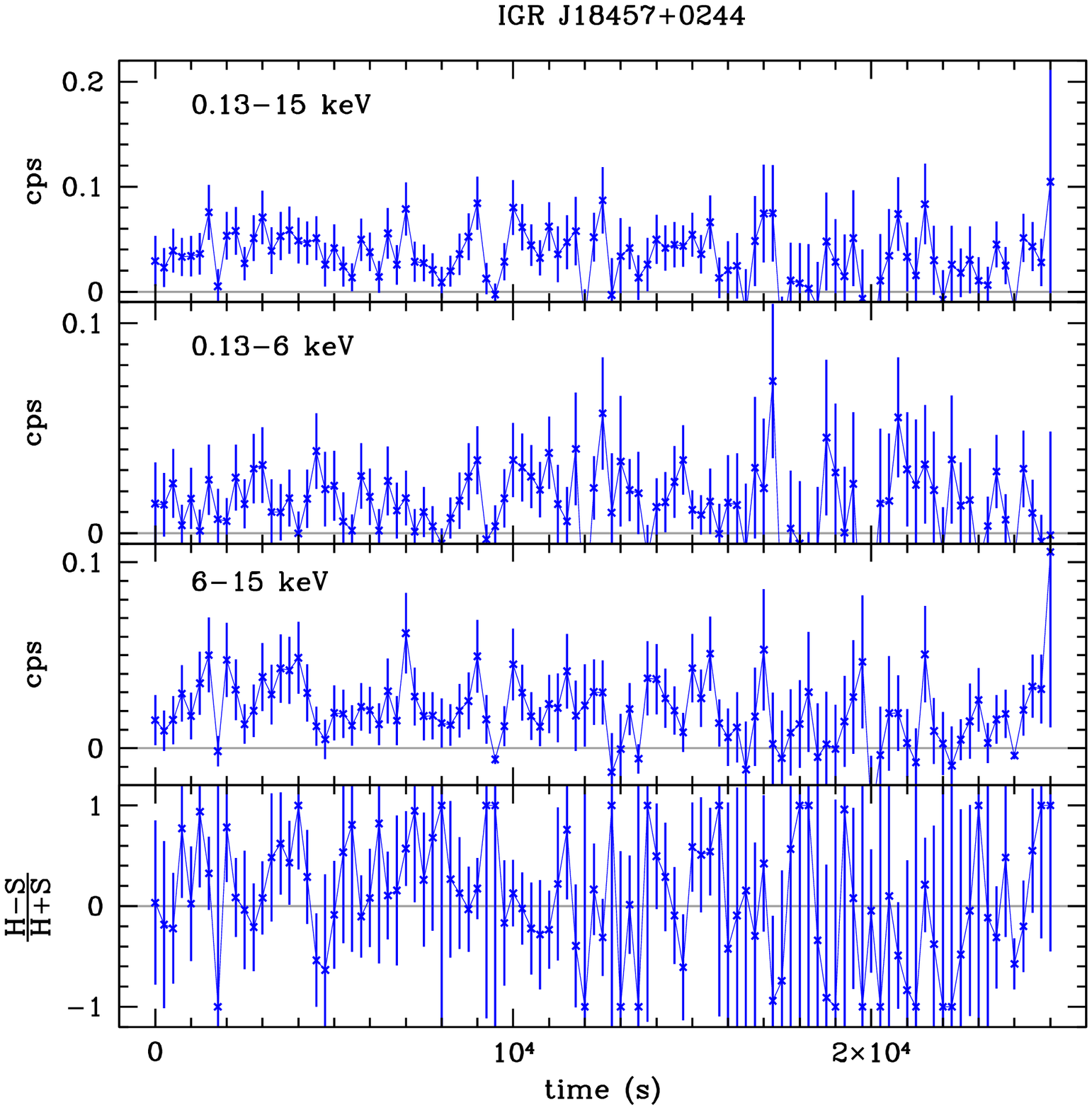}\includegraphics[width=6cm,angle=0]{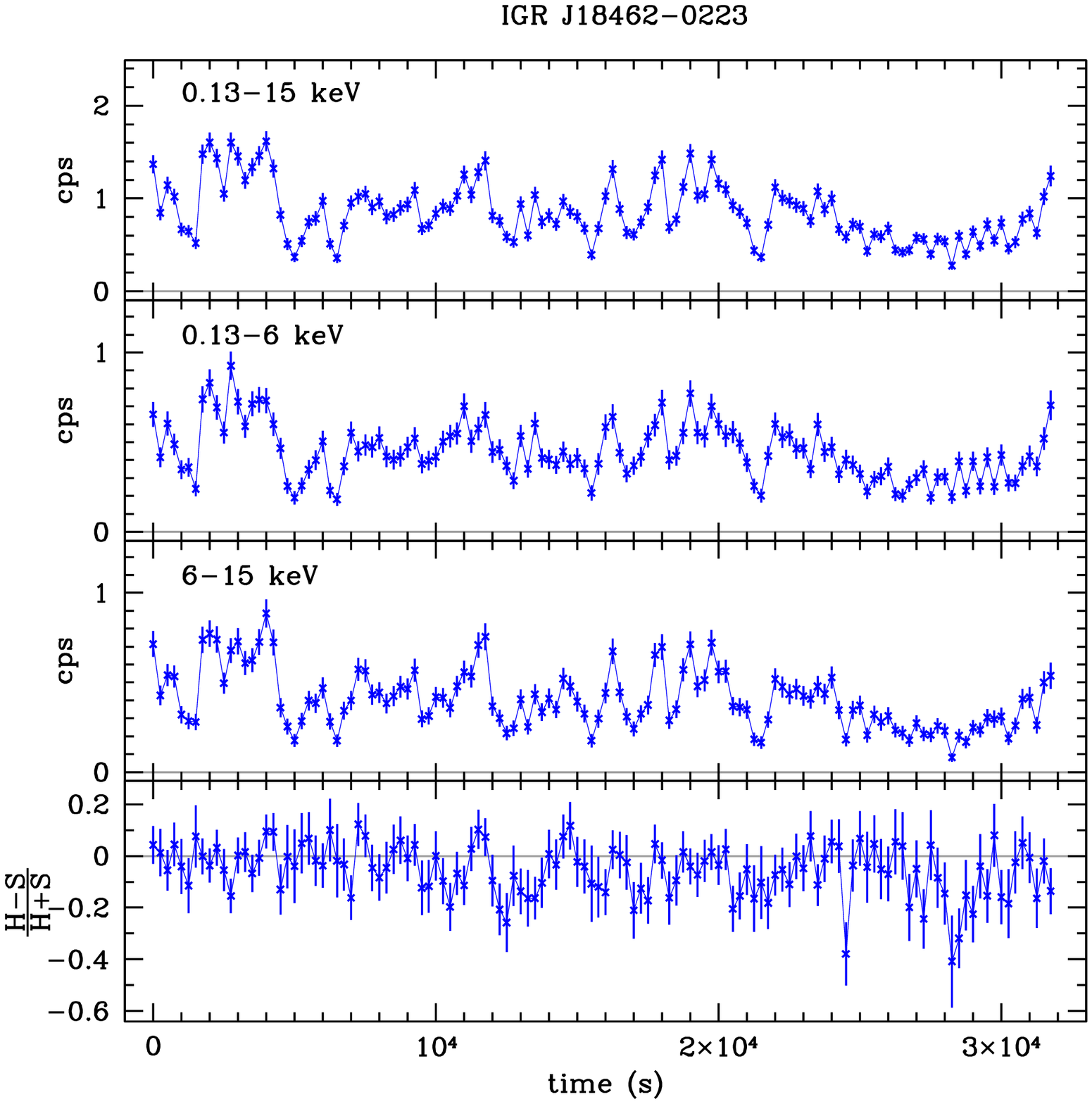}\includegraphics[width=6cm,angle=0]{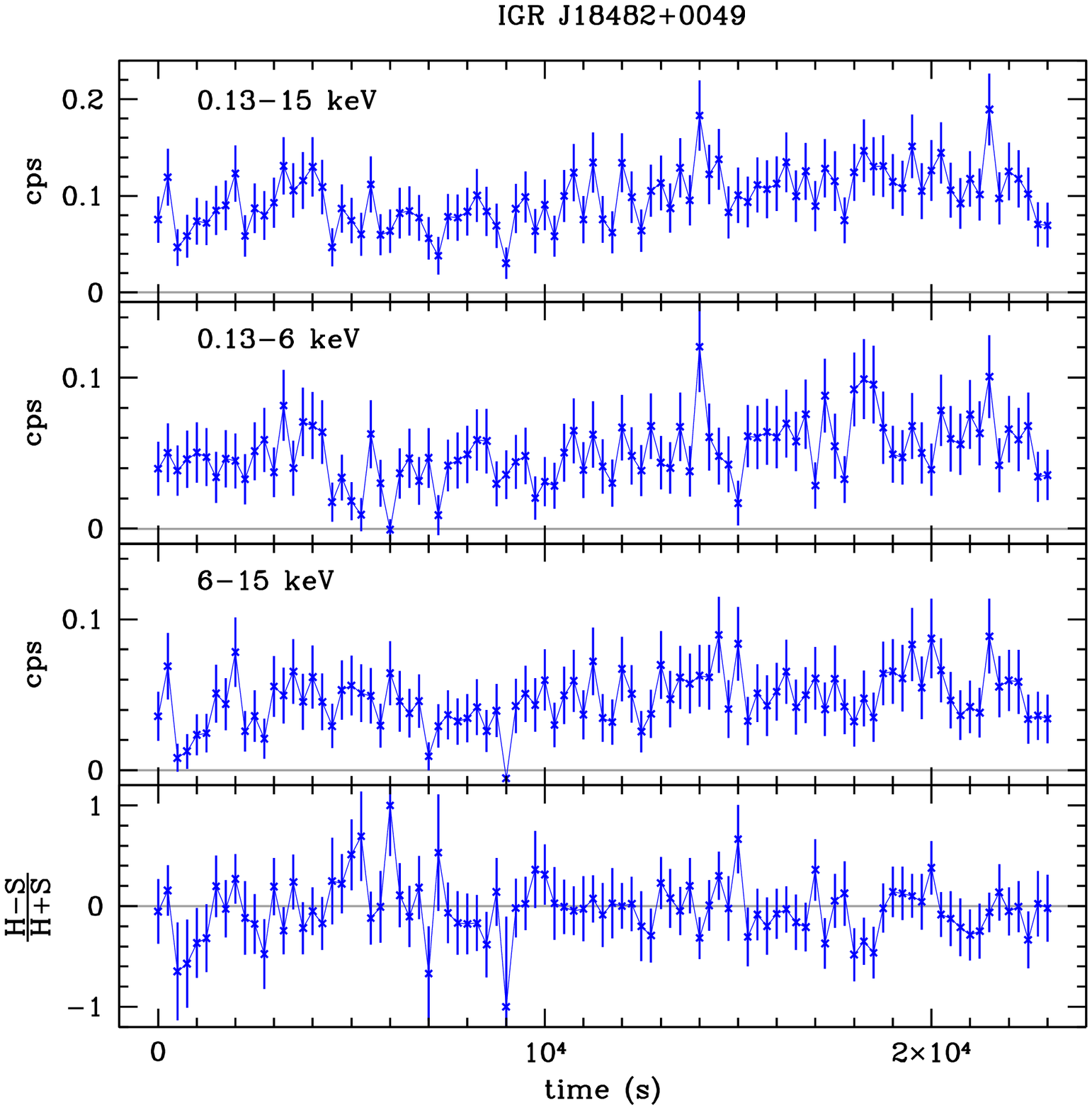}
\includegraphics[width=6cm,angle=0]{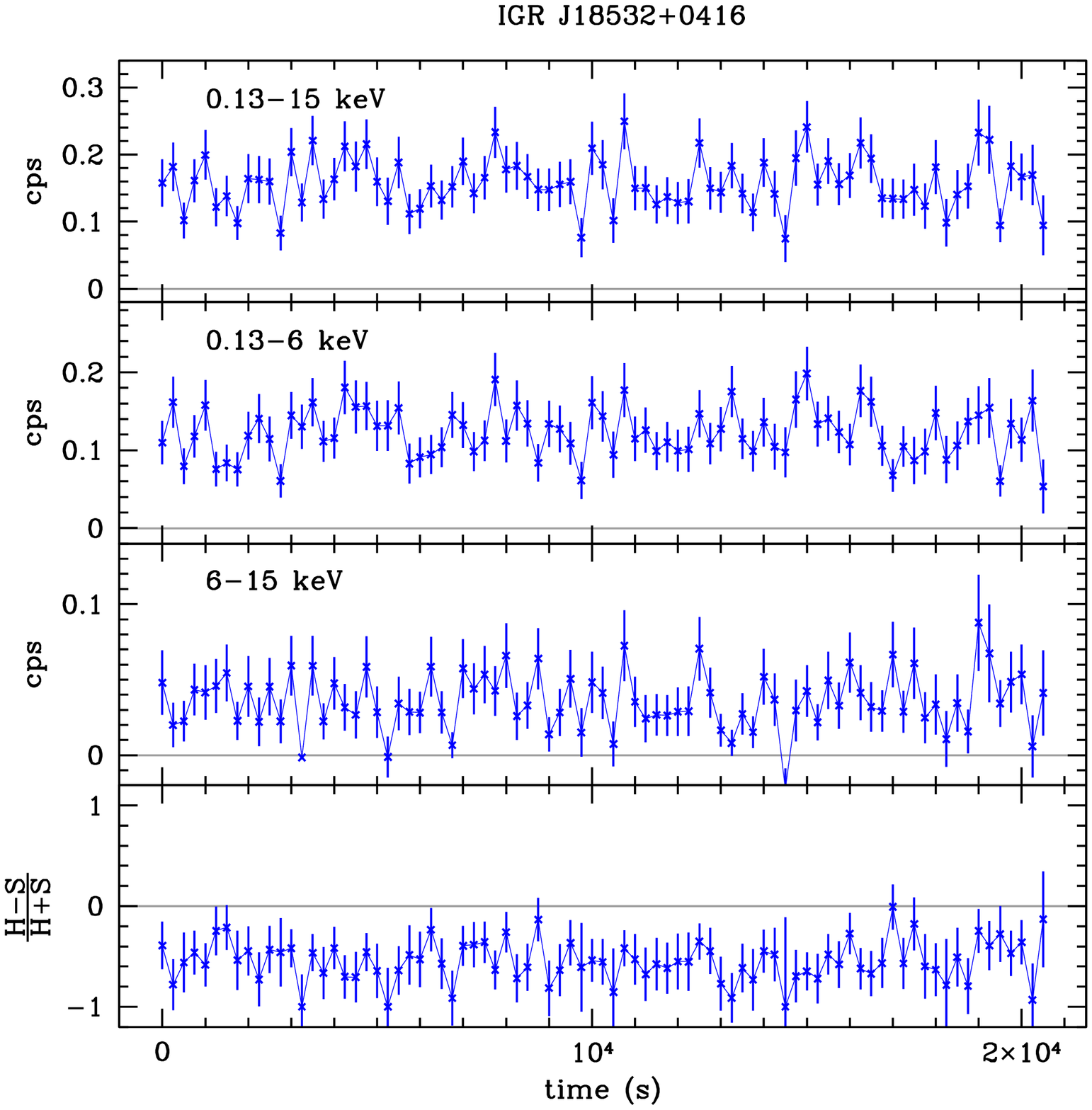}\includegraphics[width=6cm,angle=0]{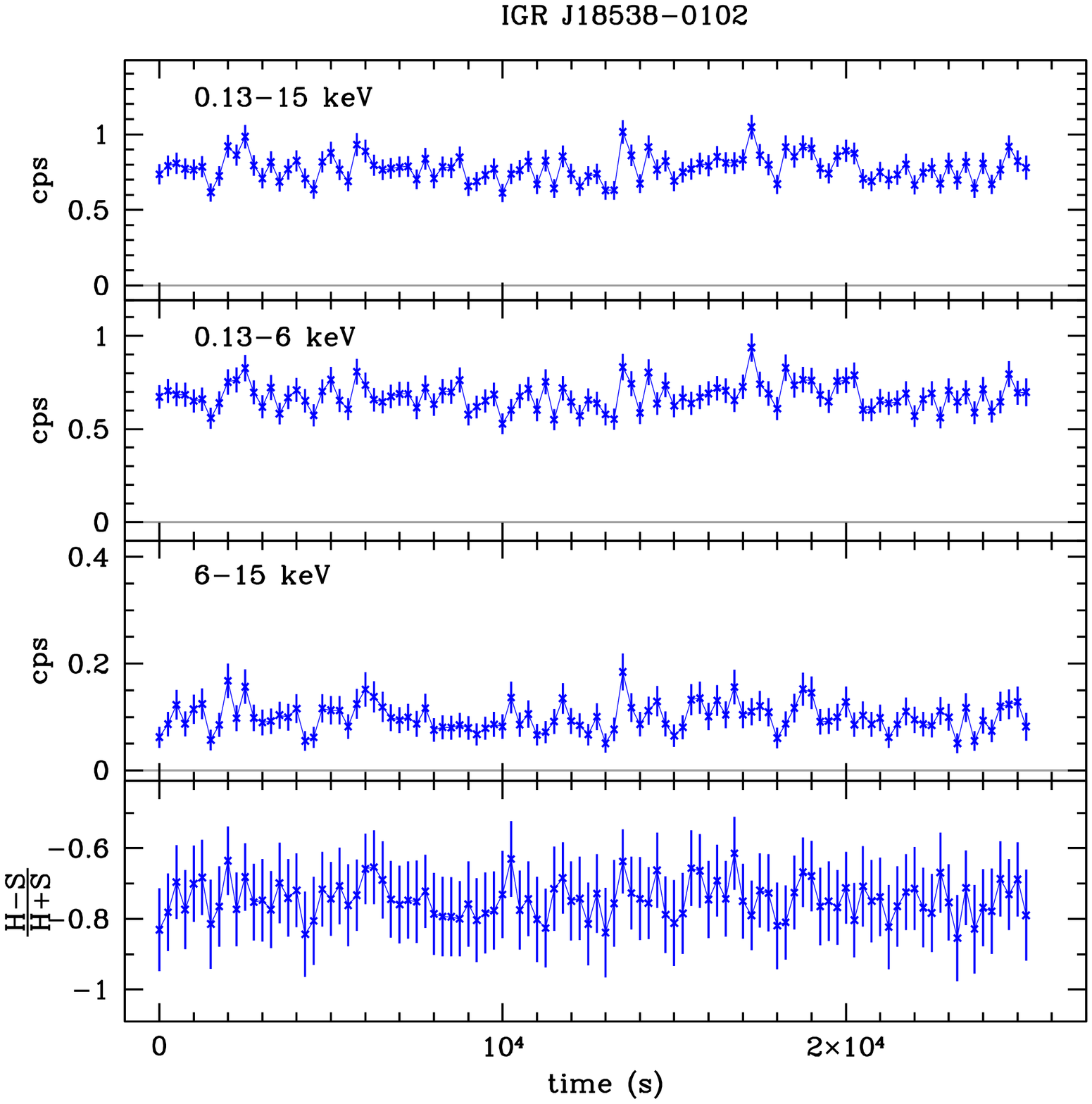}
\caption{Background-subtracted and exposure-corrected light curves of the five target sources taken with EPIC-pn in the full energy range (0.13--15\,keV), and in the soft  (S: 0.13--6\,keV) and hard bands (H: 6--15\,keV). The time resolution is 250\,s. The hardness ratio is defined as $\frac{H-S}{H+S}$.}
\label{fig_lc}
\end{figure*}

Figure\,\ref{lc_bg} presents the light curve for the particle background, as well as for the source and background extraction regions as defined above. Light curves and spectra were created for the source and background regions in 0.1--10\,keV (\texttt{PATTERN}$\le$12) for MOS1/2 (5\,s resolution), and in 0.13--15\,keV (\texttt{PATTERN}$\le$4) for pn (0.1\,s resolution). We accounted for the difference in the size of the extraction areas by running \texttt{epiclccorr} on the light curves (which  subtracted this scaled background count rate from the source count rate while correcting for exposure, PSF, and vignetting effects), and by running \texttt{backscale} on the spectra. Spectral RMF and ARF files were generated using \texttt{rmfgen} and \texttt{arfgen}, respectively. These files were grouped with the source and background spectral files by employing \texttt{specgroup} with a minimum of 20 counts per bin and a maximum oversampling factor of 3. Figures \ref{fig_lc} and \ref{fig_spec} present the light curves and spectra, respectively, of the five sources in our study. We searched for periodic signals in the raw (i.e., not corrected for the background) source light curves from pn using both \texttt{efsearch} and the fast algorithm for Lomb-Scargle periodograms developed by \citet{pre89} with error analysis from \citet{hor86}. Spectral fits employed the abundances of \citet{wil00} and the photo-ionization cross sections of \citet{bal92}. In the following section, we will discuss the results of each source in detail.

%__________________________________________________________________LC

%__________________________________________________________________
\section{Results \& Discussion}
\label{sec_res}

\subsection{IGR~J18457$+$0244}

\begin{figure*}[!t] 
\centering
\includegraphics[width=\textwidth,angle=0]{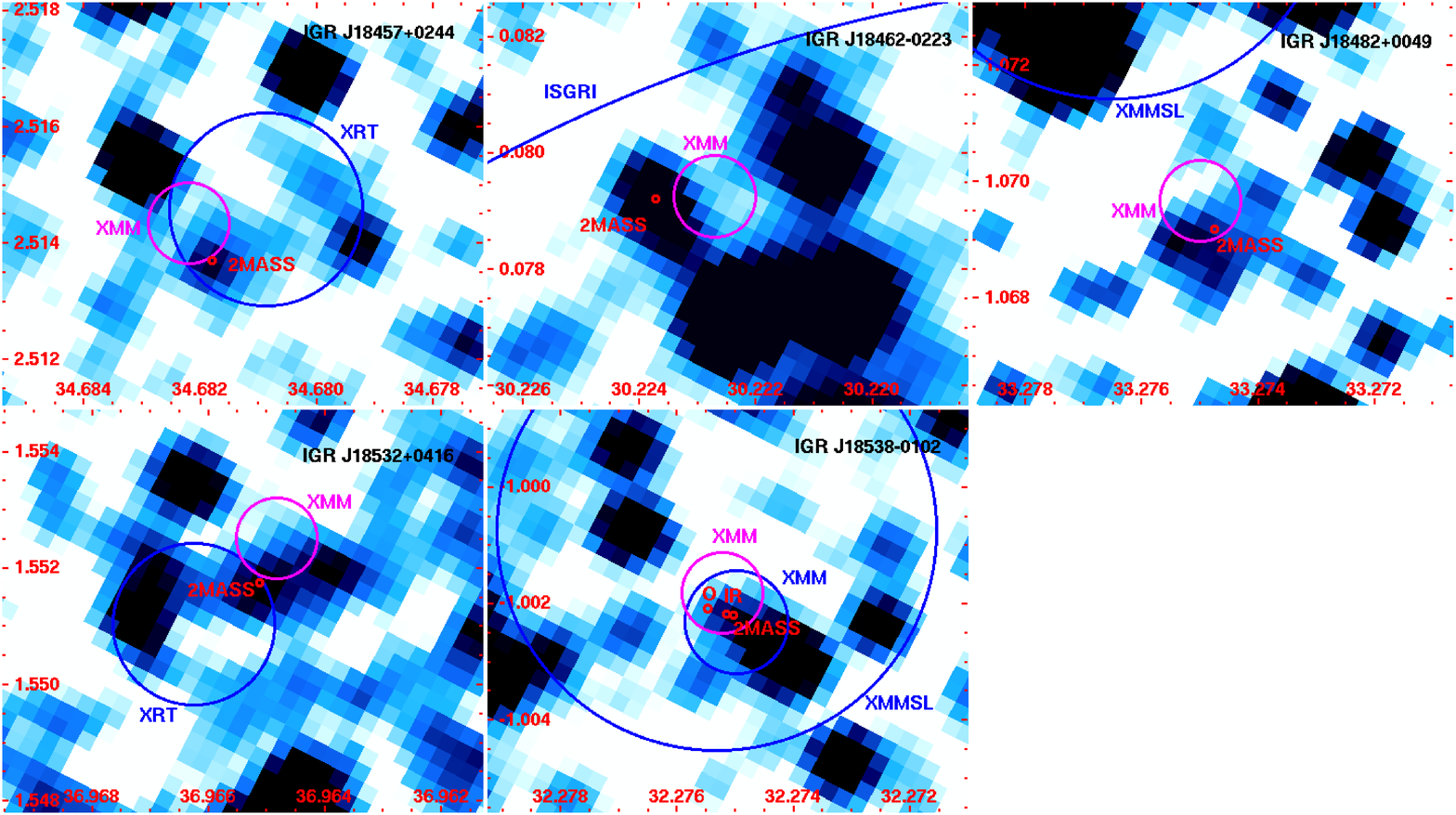}
\caption{2MASS $J$-band images of the fields of the five targets in Galactic coordinates \citep{skr06}. The positions obtained with \emph{XMM-Newton} (this work) are shown as magenta circles (2$\overset{\prime\prime}{.}$5 accuracy at 90\% confidence), while the blue and red circles represent the error circles (90\% confidence) of previously proposed X-ray and optical/infrared associations, respectively. Please see the text for details. }
\label{fig_2mass}
\end{figure*}

Soon after the discovery of this source by \citet{bir10}, it was followed up by a \swift\ observation in which a soft X-ray counterpart was detected at the 4-$\sigma$ level in the 0.3--10-keV band \citep{lan10}. The X-ray position that we obtain for \object{IGR~J18457$+$0244} is R.A. $=18^{\mathrm{h}} 45^{\mathrm{m}} 40\overset{\mathrm{s}}{.}30$ and Dec. $= +02^{\circ} 42^{\prime} 11\overset{\prime\prime}{.}2$. Our \xmm\ position is 4.9$^{\prime\prime}$ away from the \swift\ position listed in \citet{lan10}, but the latter has a reported uncertainty radius of 6$^{\prime\prime}$ so the positions are statistically compatible. The nearest infrared source in the 2MASS catalog (Fig.\,\ref{fig_2mass}), {2MASS~J18454039$+$0242088}, is 2.8$^{\prime\prime}$ away from the \xmm\ position (i.e., just outside the 90\%-confidence radius of 2$\overset{\prime\prime}{.}$5), and it has $J$, $H$, and $K$-band magnitudes of $\ge$16.2, 15.3$\pm$0.1, and 14.6$\pm$0.1, respectively \citep{skr06}. There are no catalogued objects from other wavelengths inside the \xmm\ error circle.

%__________________________________________________________________Spec
\begin{figure*}[!t]
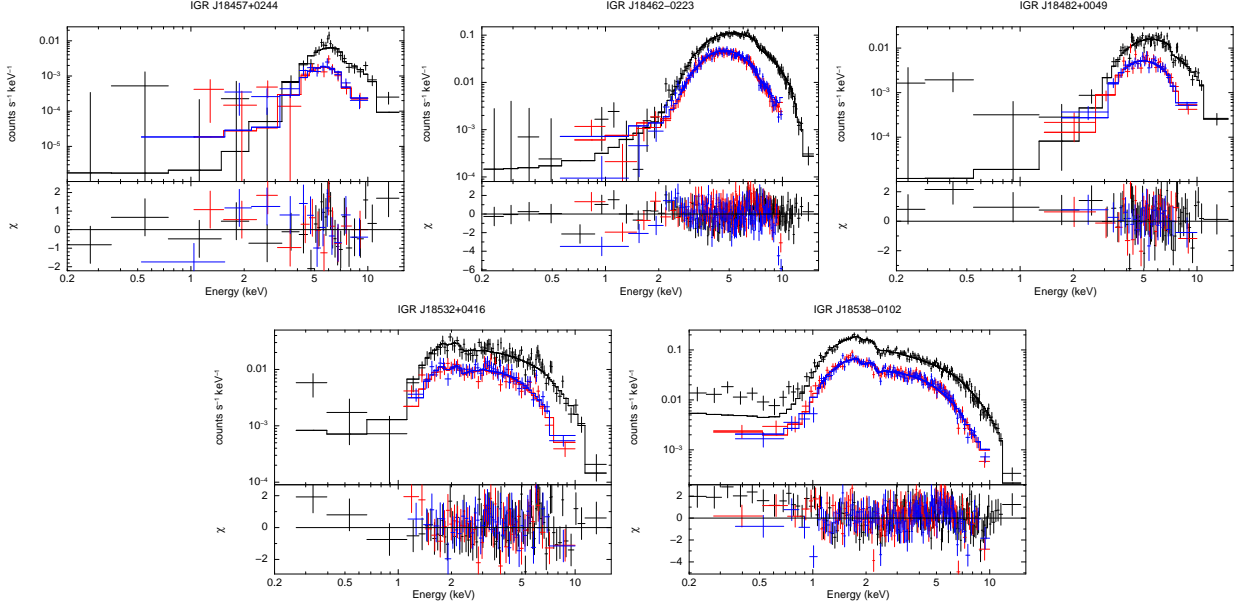
 
\centering
\includegraphics[width=4cm,angle=-90]{fig5a.ps}\hspace{2mm}\includegraphics[width=4cm,angle=-90]{fig5b.ps}\hspace{2mm}\includegraphics[width=4cm,angle=-90]{fig5c.ps}
\includegraphics[width=4cm,angle=-90]{fig5d.ps}\hspace{2mm}\includegraphics[width=4cm,angle=-90]{fig5e.ps}
\caption{Spectra (corrected for the background) in the 0.13--15-keV band of the five sources as gathered with pn (black), MOS1 (red), and MOS2 (blue). Each bin collects a minimum of 20 counts, and the spectra have been modeled with an absorbed power law whose parameters are given in Table\,\ref{tab_spec}. }
\label{fig_spec}
\end{figure*}

\begin{figure*}[!t] 
\centering
\includegraphics[width=7.5cm,angle=0]{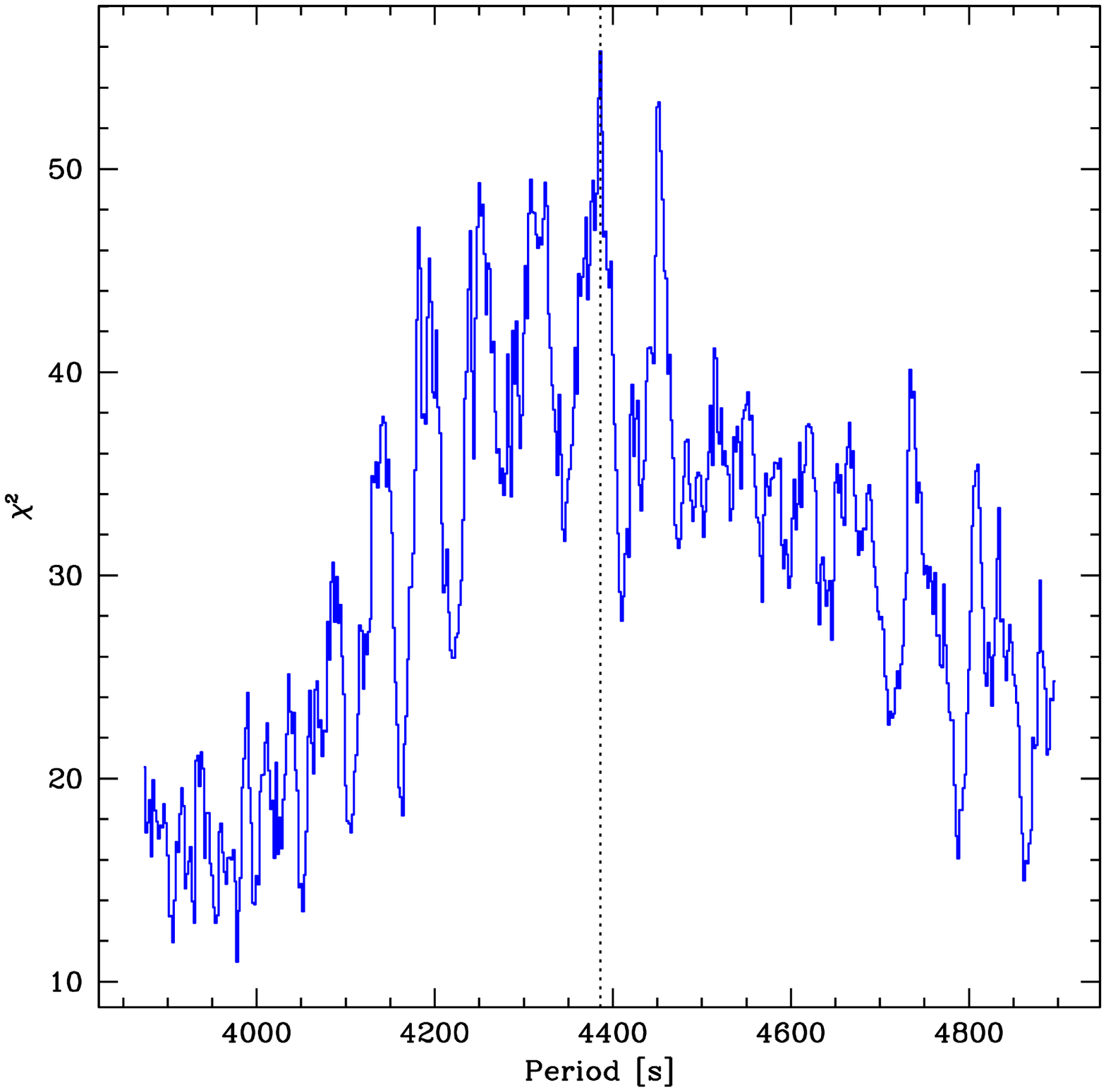}\includegraphics[width=7.5cm,angle=0]{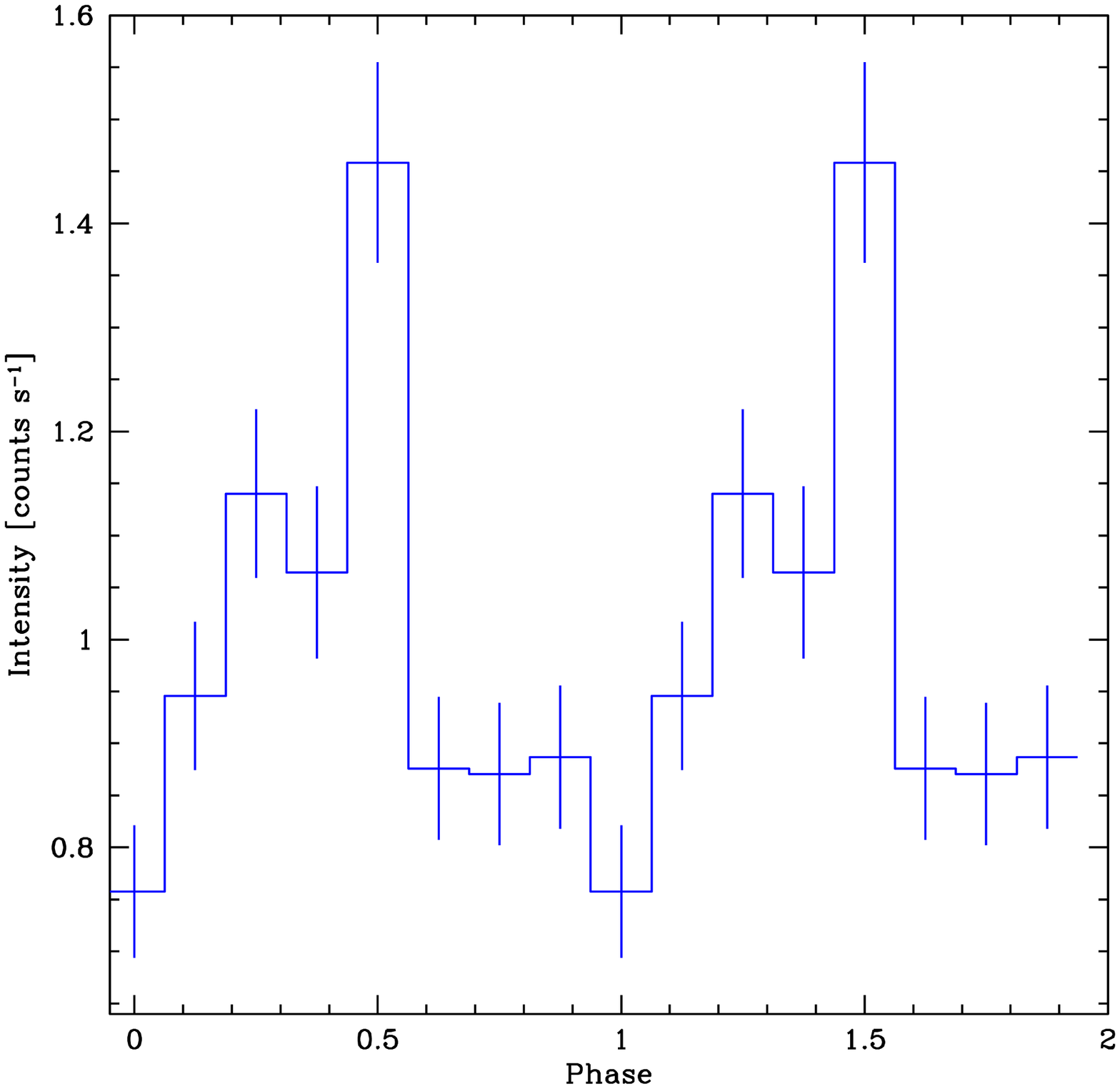}
\caption{\emph{Left}: Results from a periodicity search ($\chi^{2}$ distribution) performed on the pn light curve (0.13--15\,keV) of IGR~J18457$+$0244 centered at 4386\,s (vertical line), with 8 bins per period, and a resolution of 2\,s. \emph{Right}: Pulse profile showing two phases of IGR~J18457$+$0244 for a period of 4386\,s. The zero phase, which corresponds to the phase where the flux is at a minimum, is MJD\,55660.00609.} 
\label{fig_18457_efsearch}
\end{figure*}

Figure\,\ref{fig_lc} presents the pn light curve for \object{IGR~J18457$+$0244}. Searching for periodicities in this light curve uncovers a potential signal at a period of $4400$\,s (8 bins per trial period) with $\chi^{2} = 56$ (6-$\sigma$ significance, not corrected for the number of trials) and a pulse fraction ($\equiv \frac{I_{\mathrm{max}} - I_{\mathrm{min}}}{I_{\mathrm{max}} + I_{\mathrm{min}}}$) of 32\%$\pm$7\%. Large wings surround the main peak due to the long period of this potential signal which means that only 5 cycles are contained within the full 25-ks observation. The $\chi^{2}$ distribution and folded light curve are presented in Fig.\,\ref{fig_18457_efsearch}. This signal is only seen with \texttt{efsearch}, and not with the Lomb-Scargle periodogram nor in the power spectrum. Using the Leahy-normalized power spectrum \citep{Lea83}, we determined a 90\% confidence upper limit of 25.33 on the Leahy Power \citep{vdk89} which converts to an upper limit of 14\% on the fractional r.m.s. expected for a periodic signal due to noise (i.e., less than the pulse fraction that we measured).

%__________________________________________________________________Spectral Parameters
%
\begin{deluxetable}{ l l c c c c c c c c c c c }
\tabletypesize{\scriptsize}
\rotate
\tablecolumns{13} 
\tablewidth{0pc} 
\tablecaption{Spectral parameters from models fit to the pn and MOS spectra of the five targets. \label{tab_spec}}
\tablehead{
\colhead{Source Name}					& 
\colhead{\texttt{Xspec} Model}			& 
\colhead{$C$}						& 
\colhead{$N_\mathrm{H}$}				& 
\colhead{$\Gamma$ or $\tau$}			& 
\colhead{$E_{\mathrm{cut}}$ or $T_{0}$}	&
\colhead{$kT$}						& 
\colhead{$E\left(\mathrm{K}\alpha\right)$}	&   
\colhead{$EW\left(\mathrm{K}\alpha\right)$}	&     
\colhead{Norm. $\left(\mathrm{K}\alpha\right)$}	&
\colhead{$^{\mathrm{unabs}}F_{\mathrm{0.5-10 keV}}$}		&        		
\colhead{$\chi_{\nu}^{2}$/dof}				\\    	
\colhead{}								& 	
\colhead{}								& 	
\colhead{}								& 
\colhead{$10^{22}$\,cm$^{-2}$}				& 
\colhead{}								&
\colhead{keV}							&
\colhead{keV}							&
\colhead{keV}							&
\colhead{eV}								&
\colhead{$10^{-6}$\,photons\,cm$^{-2}$\,s$^{-1}$}						&
\colhead{$10^{-12}$\,erg\,cm$^{-2}$\,s$^{-1}$}	&
\colhead{}   			
}

\startdata
\vspace{1mm}

IGR~J18457$+$0244 & c*phabs*pow				& $1.04_{-0.15}^{+0.17}$	& $84_{-16}^{+20}$	& 2.3$\pm$0.7	& 	& 	& 	& 	& 	& 7.7$\pm$0.5	 & 1.13/49		\\
\vspace{1mm}
& c*phabs*(pow+gauss)		& $1.02_{-0.14}^{+0.17}$	& $74_{-17}^{+19}$	& 1.9$\pm$0.7	& 	&  & $6.01_{-0.06}^{+0.04}$ & 300$\pm$150	& $8_{-3}^{+4}$ 	& 4.1$\pm$0.2 & 0.85/46	\\
\vspace{1mm}
& c*phabs*(cutoffpl+gauss)		& $1.02_{-0.15}^{+0.16}$	& $72_{-16}^{+15}$	& $1.8_{-1.3}^{+0.9}$	& $\ge$0.01	& 	& $6.01_{-0.06}^{+0.04}$	& 310$\pm$160 & $8_{-3}^{+4}$ & 4$\pm$3 & 0.87/45	\\
\vspace{1mm}
& c*phabs*(bbodyrad+gauss)		& $1.03_{-0.15}^{+0.17}$	& $57_{-14}^{+15}$	& 	& 	& $2.1_{-0.4}^{+0.6}$	& $6.01_{-0.06}^{+0.05}$	& 310$\pm$160 & $6_{-2}^{+3}$ & 1.4$\pm$0.2 & 0.92/46	\\
\vspace{1mm}
& c*phabs*(comptt+gauss)		& $1.02_{-0.15}^{+0.17}$	& $72_{-30}^{+19}$	& $\le200$	& $\le$2.8 & $\le$115	& $6.01_{-0.06}^{+0.04}$	& 300$\pm$150 & $8_{-3}^{+4}$	& 2.2$\pm$0.3 & 0.89/44	\\
\hline
\vspace{1mm}

IGR~J18462$-$0223	 & c*phabs*pow				& 1.13$\pm$0.03		& 28$\pm$1		& 1.5$\pm$0.1	& 	& 	& 	& 	& 	& 36.5$\pm$0.7	& 1.32/337		\\
\vspace{1mm}
& c*phabs*(pow+gauss)		& 1.13$\pm$0.03		& 28$\pm$1		& 1.5$\pm$0.1	& 	&  & $6.41_{-0.04}^{+0.05}$ & 70$\pm$32 & 15$\pm$5 & 27.7$\pm$0.6 & 1.25/334	\\
\vspace{1mm}
& c*phabs*(cutoffpl+gauss)		& 1.13$\pm$0.03		& 21$\pm$2		& $-$0.8$\pm$0.6	& $3.3_{-0.7}^{+1.3}$	& 	& 6.41$\pm$0.05	& $48_{-27}^{+43}$ & $10.1_{-0.4}^{+0.5}$	& 16$\pm$10 & 1.14/333	\\
\vspace{1mm}
& c*phabs*(bbodyrad+gauss)		& 1.13$\pm$0.03	& 19$\pm$1	& 	& 	& $2.17_{-0.05}^{+0.06}$	& $6.40_{-0.05}^{+0.06}$	& $41_{-23}^{+16}$ & $8_{-4}^{+5}$ & 14.9$\pm$0.3 & 1.16/334	\\
\vspace{1mm}
& c*phabs*(comptt+gauss)		& 1.13$\pm$0.03	& $21_{-3}^{+4}$	& $12_{-6}^{+5}$	& $\le$1.7 & $\le$229	& $6.41_{-0.06}^{+0.05}$	& $49_{-17}^{+44}$ & 10.3$\pm$0.5	& 16$\pm$14 & 1.15/332	\\
\hline
\vspace{1mm}
IGR~J18482$+$0049 & c*phabs*pow				& $1.03_{-0.07}^{+0.08}$		& 44$\pm$5		& 2.0$\pm$0.3	& 	& 	& 	& 	& 	& 8$\pm$2 & 1.01/108		\\
\vspace{1mm}
& c*phabs*bbodyrad			& $1.04_{-0.07}^{+0.08}$		& $31_{-3}^{+4}$	& 	& 	& $1.9_{-0.1}^{+0.2}$	& 	&  & 	& 2.4$\pm$0.4 & 1.05/108	\\
\hline
\vspace{1mm}

IGR~J18532$+$0416 & c*phabs*pow				& 1.16$\pm$0.07		& $3.3_{-0.3}^{+0.4}$		& 1.4$\pm$0.1	& 	& 	& 	& 	& 	& 1.8$\pm$0.1	 & 1.05/166		\\
\vspace{1mm}
& c*phabs*(pow+gauss)							& 1.16$\pm$0.07		& 3.4$\pm$0.4		& 1.4$\pm$0.1	& 	&  & $6.10_{-0.04}^{+0.06}$ & $253_{-109}^{+122}$ & $4_{-1}^{+2}$	& 1.8$\pm$0.1 & 0.94/163	\\
\vspace{1mm}
& c*phabs*(cutoffpl+gauss)		& 1.16$\pm$0.07		& 3.1$\pm$0.6		& $1.1_{-0.6}^{+0.4}$	& $\ge$6	& 	& $6.10_{-0.04}^{+0.06}$	& 250$\pm$90 & $4_{-1}^{+2}$	& 1.7$\pm$0.2 & 0.94/162	\\
\vspace{1mm}
& c*phabs*(bbodyrad+gauss)		& 1.16$\pm$0.07	& 1.2$\pm$0.2	& 	& 	& 1.53$\pm$0.08	& $6.09_{-0.04}^{+0.06}$	& $207_{-77}^{+96}$ & $4_{-1}^{+2}$	& 1.2$\pm$0.1 & 1.23/163	\\
\vspace{1mm}
& c*phabs*(comptt+gauss)		& $1.17_{-0.07}^{+0.08}$	& $2.1_{-0.5}^{+1.6}$	& $1.5_{-1.2}^{+3.6}$	& $\le$0.8 & $\le$294	& $6.11_{-0.05}^{+0.06}$	& 270$\pm$110 & $4_{-1}^{+2}$	& $\le$6.5 & 0.94/161 	\\
\hline
\vspace{1mm}

IGR~J18538$-$0102	 & c*phabs*pow				& $1.02_{-0.02}^{+0.03}$		& 1.98$\pm$0.08		& 1.57$\pm$0.04	& 	& 	& 	& 	& 	& 6.3$\pm$0.1	 & 1.33/359		\\
\vspace{1mm}
& c*phabs*cutoffpl			& $1.02_{-0.02}^{+0.03}$		& 1.7$\pm$0.1			& 1.1$\pm$0.2		& $9_{-3}^{+5}$ & 	& 	&  & 	& 5.6$\pm$0.8 & 1.28/358	\\

\enddata
\vspace{-5mm}
\tablecomments{Errors are quoted at 90\% confidence. $C$ is an instrumental cross-calibration coefficient which is fixed at 1 for pn and variable for MOS. In the \texttt{Xspec} formalism: phabs $=$ photoelectric absorption; pow $=$ power law; cutoffpl $=$ cutoff power law; bbodyrad $=$ radial blackbody; comptt $=$ Compton thermalization; and gauss $=$ Gaussian.}
\end{deluxetable}

We note that aperiodic variability on a $\sim$4-ks timescale (i.e., a large fraction of the observation duration) would also lead to multiple wings surrounding a central peak in the $\chi^{2}$ distribution, so this can not be excluded. Thus, an intriguing possibility is that the weak periodicity at 4.4\,ks is a quasi-periodic oscillation (QPO) related to the motion of material along the innermost stable orbit of an accretion disk. Such QPOs are a common feature of the power spectrum of black hole binaries. They have been confirmed in one AGN thus far (\object{RE~J1034+396}) where the peak frequency of the QPO is $2.7\times10^{-4}$\,Hz (i.e., 3.7-ks periodicity) \citep{gie08}, and they have been proposed in other AGN as well \citep[e.g.,][]{esp08,gup09,lac09}. These frequencies are similar in scale to the 4.4-ks periodicity that we see in \object{IGR~J18457$+$0244}. 

A power law fit to the X-ray spectrum (Fig.\,\ref{fig_spec}) yields a column density (\nh\ $= (84_{-16}^{+20})\times10^{22}$\,\cmsq) that is larger than that expected along the line of sight \citep{kal05}, which suggests that it is intrinsic to the system, and a steep photon index ($\Gamma =$ 2.3$\pm$0.7). These parameters are in good agreement with those from \swift-XRT \citep{lan10}, despite the fact that our observed (i.e., not corrected for absorption) 0.5--10-keV flux is $4.6\times10^{-13}$\,\ergcms\ ($= 5.5\times10^{33} \left [ \frac{d}{10\textrm{ kpc}} \right ]^{2}$\,\ergs), i.e., lower than that of the 2--10-keV flux observed with XRT.  The quality of the fit is good ($\chi_{\nu}^{2}$/dof $=$ 1.13/49), but a few residuals around 6\,keV hint at an iron fluorescence line. Adding a Gaussian to the model reduces the $\chi_{\nu}^{2}$/dof to 0.85/46. The line energy is $6.01_{-0.04}^{+0.06}$\,keV with an equivalent width ($EW$) of 300$\pm$150\,eV. A good fit ($\chi_{\nu}^{2}$/dof $=$ 0.92/46) is also obtained with an absorbed blackbody model that includes a Gaussian profile for the possible iron line. In this case, the blackbody temperature is $2.1_{-0.4}^{+0.6}$\,keV, and \nh\ $= (54_{-14}^{+15})\times10^{22}$\,\cmsq. Modeling the spectrum with a cutoff power law or a thermal Comptonization model (\texttt{comptt} in \texttt{Xspec}) returned upper limits on the spectral parameters. These results are summarized in Table\,\ref{tab_spec}.

%__________________________________________________________________IMG:2MASS

Assuming that the rest energy of the iron $K\alpha$ line is 6.41\,keV, this would place \object{IGR~J18457$+$0244} at a redshift of $z = 0.07(1)$, suggesting an active galactic nucleus (AGN) most likely of the Seyfert-2 class given its large absorbing column. The corresponding (unabsorbed) X-ray luminosity in the 0.5--10-keV band is $(9_{-2}^{+3})\times10^{43}$\,\ergs, assuming $H_{\circ} = 70$\,km\,s$^{-1}$\,Mpc$^{-1}$ and a flat cosmology. We point out that other than the high \nh\ value (which can also be found among obscured X-ray pulsars), the main evidence supporting the AGN designation is the iron line energy, and this detection is marginal since only 2 spectral bins in pn show significant deviations ($\ge 3\sigma$) from the continuum (1 bin for each MOS detector). On the other hand, if the potential periodic modulation can be reproduced in other observations, and if the signal were found to be coherent, it would point instead to a neutron star spin period, and the AGN would no longer be a viable explanation. 

In any case, additional long-duration observations are needed to elucidate the nature of \object{IGR~J18457$+$0244}.

\subsection{IGR~J18462$-$0223}

\citet{gre07a} discovered \object{IGR~J18462$-$0223} during a flare that lasted a few hours. The source was actually in outburst a year earlier but this flare had gone unnoticed until reexamination of archival data \citep{gre10}. A day after its discovery, it could no longer be detected. The source erupted again in 2010, which led \citet{gre10} to suggest that the source belonged to a newly-recognized class of SGXB: a supergiant fast X-ray transient (SFXT). These are HMXBs with early-type supergiant donors feeding an X-ray source that, unlike in ``classical'' SGXBs, features huge variability in its emission (luminosity swings of $\sim10^{4}$--$10^{5}$ are typical).

%__________________________________________________________________LC EFSEARCH

For \object{IGR~J18462$-$0223}, we obtain with \xmm\ a position of R.A. $=18^{\mathrm{h}} 46^{\mathrm{m}} 12\overset{\mathrm{s}}{.}68$ and Dec. $= -02^{\circ} 22^{\prime} 29\overset{\prime\prime}{.}3$ which is $1\overset{\prime}{.}5$ from (and still compatible with) the ISGRI position of \citet{gre10} which has an error radius of $1\overset{\prime}{.}6$. The 2MASS image of the field (Fig.\,\ref{fig_2mass}) shows that the nearest catalogued infrared source is located outside the \xmm\ error radius ($3\overset{\prime\prime}{.}4$ from the EPIC position): \object{2MASS~J18461279$-$0222261} with magnitudes of 14.5$\pm$0.1, 13.8$\pm$0.1, and $\ge$12.7, respectively, in the $J$, $H$, and $K$ bands \citep{skr06}. This object is also listed as \object{USNO~B-1.0~0876-0579765} \citep{mon03}, \object{DENIS~J184612.8$-$022226} \citep{den05}, and \object{GLIMPSE~G030.2231$+$00.0791} \citep{ben03}. No other objects from other wavelengths are located inside the \xmm\ error circle.

In the X-ray spectrum of \object{IGR~J18462$-$0223} shown in Fig.\,\ref{fig_spec}, we detect a large column density (\nh\ $\sim$ (2--3)$\times10^{23}$\,\cmsq) and an iron $K\alpha$ line at $6.41_{-0.04}^{+0.05}$\,keV ($EW = $70$\pm$32\,eV). The addition of a Gaussian to the power law (to account for the $K\alpha$ line) leads to a small improvement in the quality of the fit from $\chi_{\nu}^{2}$/dof $=$ 1.32/337 to 1.25/334. The best fit ($\chi_{\nu}^{2}$/dof $=$ 1.14/333) is obtained with the addition of a cutoff at $3.3_{-0.7}^{+1.3}$\,keV. The resulting photon index and column density are $\Gamma = -0.8\pm0.6$ and \nh\ $= (21\pm2)\times10^{22}$\,\cmsq, respectively. A blackbody (\texttt{bbodyrad}) also provides a good fit to the X-ray spectra ($\chi_{\nu}^{2}$/dof $=$1.16/334). For a Compton thermalization model (\texttt{compTT}), the plasma temperature and energy of the seed photons could not be constrained.

The spectral shape is reminiscent of other wind-accreting X-ray binaries discovered by \integ, particularly that of the obscured SGXB pulsar in the Norma Arm \object{IGR~J16393$-$4643} \citep{bod06}: both sources have large \nh\ values, hard power law continua, iron line(s), and low-energy cutoffs. It is important to note that the cutoff energy in \object{IGR~J18462$-$0223} is below 20\,keV. This cutoff is required since the photon index in the ISGRI band (20--60\,keV) is $\Gamma=2.5\pm0.3$ \citep{gre10}, i.e., steeper than the slope of 1.5$\pm$0.1 that we measure for an absorbed power law between 0.5 and 10\,keV. In HMXB systems, such low cutoff energies are typical of accreting neutron stars \citep{nag89}.

%__________________________________________________________________LC EFSEARCH
\begin{figure*}[!t] 
\centering
\includegraphics[width=7.5cm,angle=0]{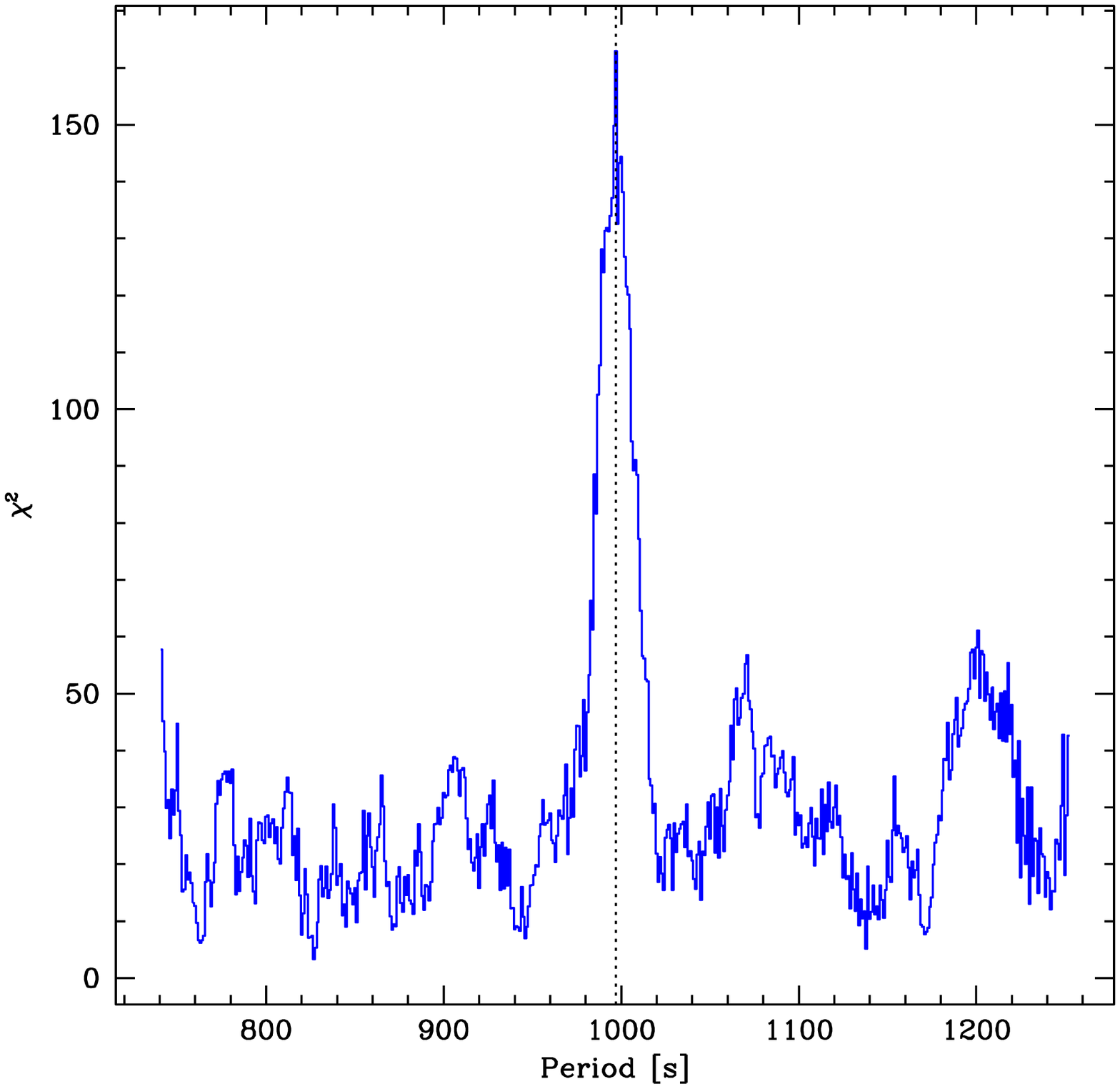}\includegraphics[width=7.5cm,angle=0]{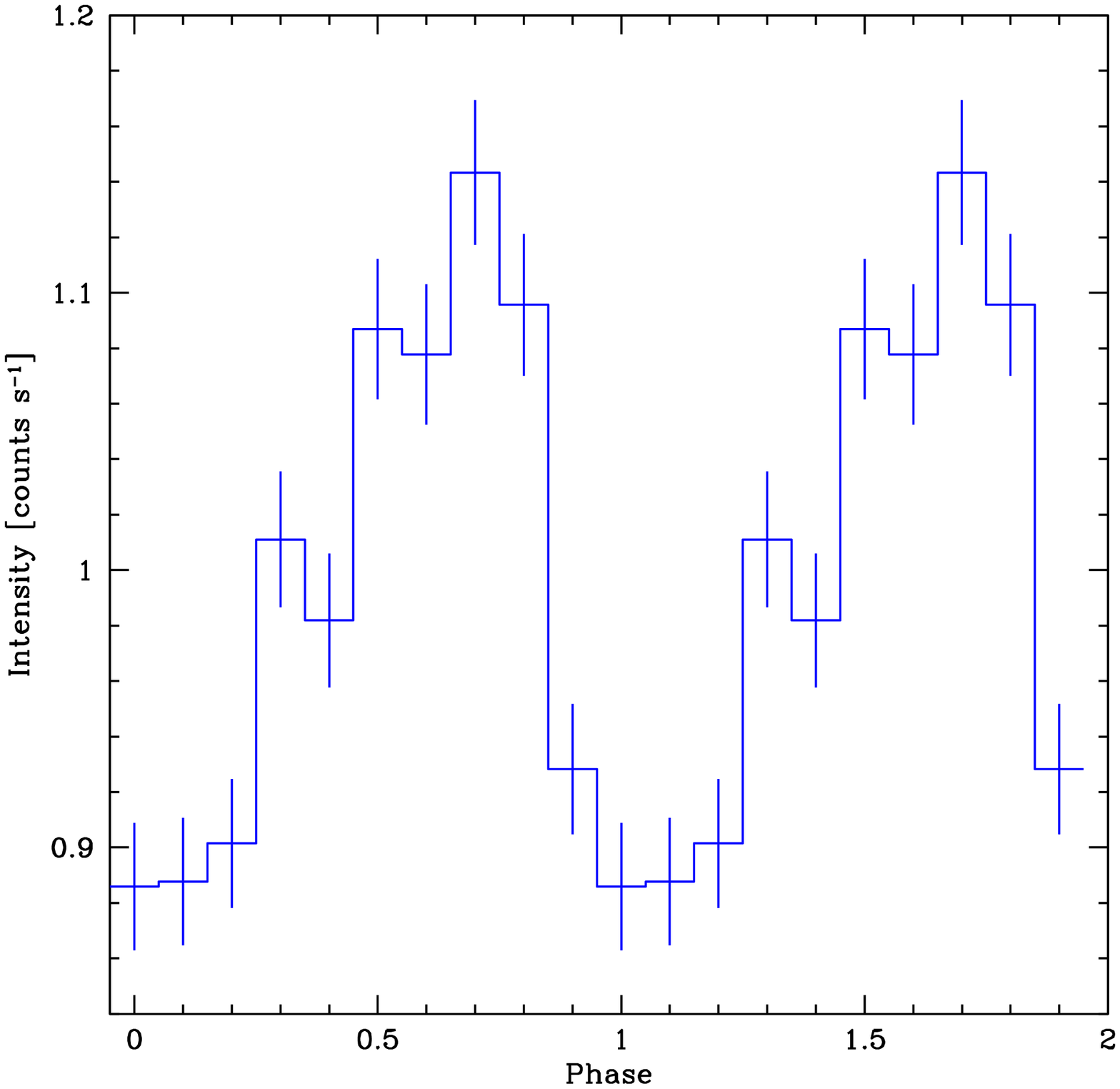}
\caption{\emph{Left}: Results from a periodicity search ($\chi^{2}$ distribution) performed on the pn light curve (0.13--15\,keV) of IGR~J18462$-$0223 centered at 997\,s (vertical line), with 10 bins per period, and a resolution of 1\,s. \emph{Right}: Pulse profile showing two phases of IGR~J18462$-$0223 for a period of 997\,s. The zero phase (phase in which the flux is at a minimum) corresponds to MJD\,55669.00808.}
\label{lc_efsearch}
\end{figure*}

Therefore, we searched for periodic modulations in the pn light curve of \object{IGR~J18462$-$0223} and we found a coherent pulsation at a period of 997$\pm$1\,s with $\chi^{2}=162$ for 10 bins per trial period ($\sim$12$\sigma$ significance, not corrected for the number of trials). The pulse profile folded on a period of 997\,s and beginning at MJD\,55669.00808(1) is shown in Fig.\,\ref{lc_efsearch}. The pulse fraction is 12\%$\pm$2\%. Thus, the compact object hosted by \object{IGR~J18462$-$0223} is an accreting neutron star whose magnetic and spin axes are misaligned.

The SFXT designation for \object{IGR~J18462$-$0223} rests on its sporadic detection history in the hard X-rays \citep{gre10}. In our 30-ks long and continuous observation (Fig.\,\ref{fig_lc}), we find no evidence of the large variability associated with this class of objects. The dynamic range for the 0.13--15-keV flux is an order of magnitude or less. This is more typical of persistently-emitting SGXB systems. Assuming that the source resides within the Scutum Arm whose tangent is $\sim$7\,kpc away \citep[e.g.,][]{rus03}, its unabsorbed flux (0.5--10 keV) of $1.6\times10^{-11}$\,\ergcms\ converts to a luminosity of around $10^{35}$\,\ergs. This is a typical luminosity for an SFXT in the active state \citep[e.g.,][]{rom11}, but it is an order of a magnitude lower than in other persistently-emitting SGXBs \citep[e.g.,][]{wal06}, unless the source were located more than 7\,kpc from us.

Overall, the timing and spectral characteristics of \object{IGR~J18462$-$0223} suggest that this system is an obscured (and probably distant) SGXB pulsar, and one of the few examples of a highly-obscured HMXB in the direction of the Scutum Arm. The source could represent an intermediate SGXB-SFXT system like \object{IGR~J16479$-$4514}, which would make \object{IGR~J18462$-$0223} the seventh (out of $\sim$20 known SFXT candidates) to have a measured spin period.

%__________________________________________________________________LC EFSEARCH
\begin{figure*}[!t] 
\centering
\includegraphics[width=7.5cm,angle=0]{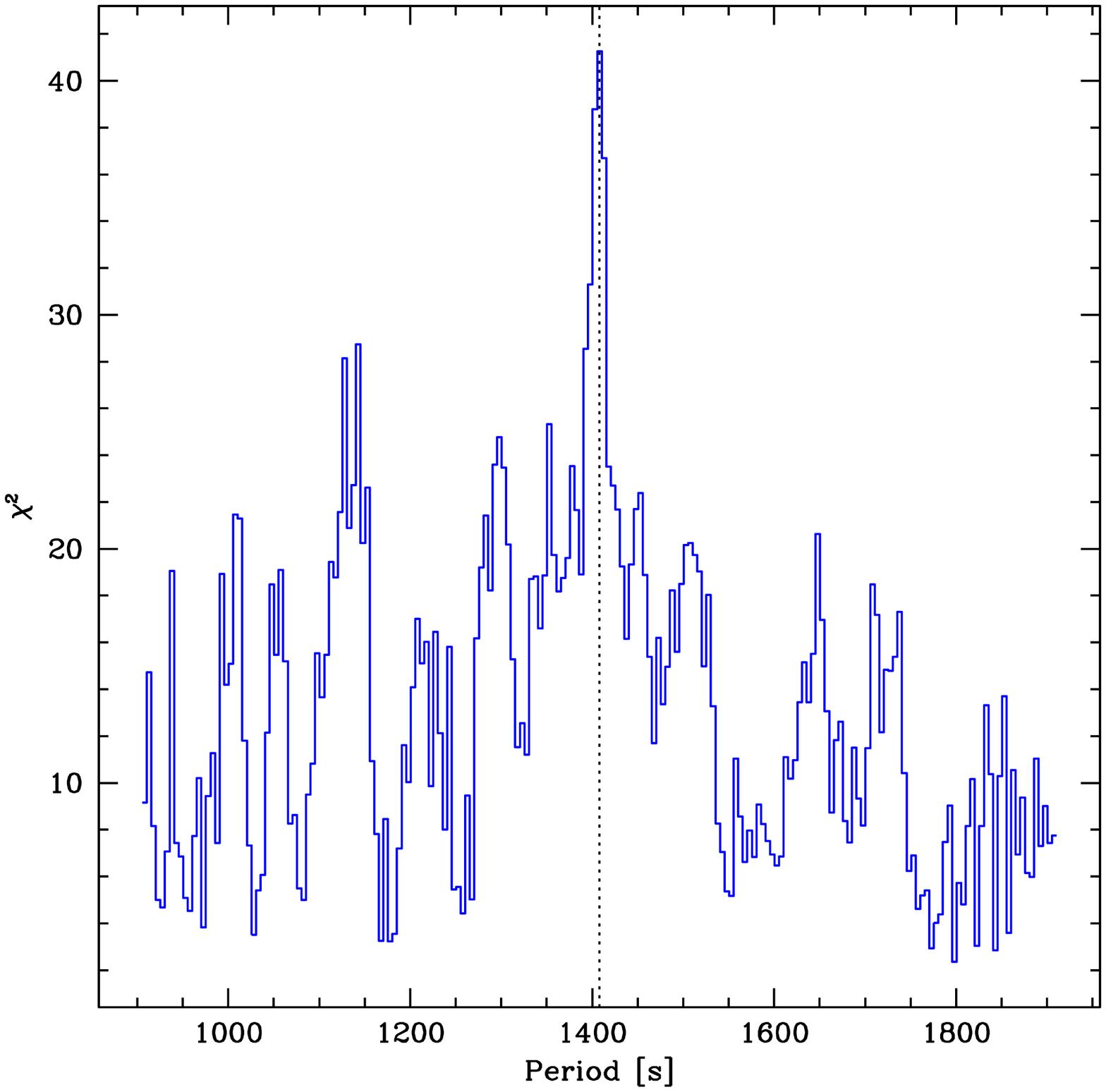}\includegraphics[width=7.5cm,angle=0]{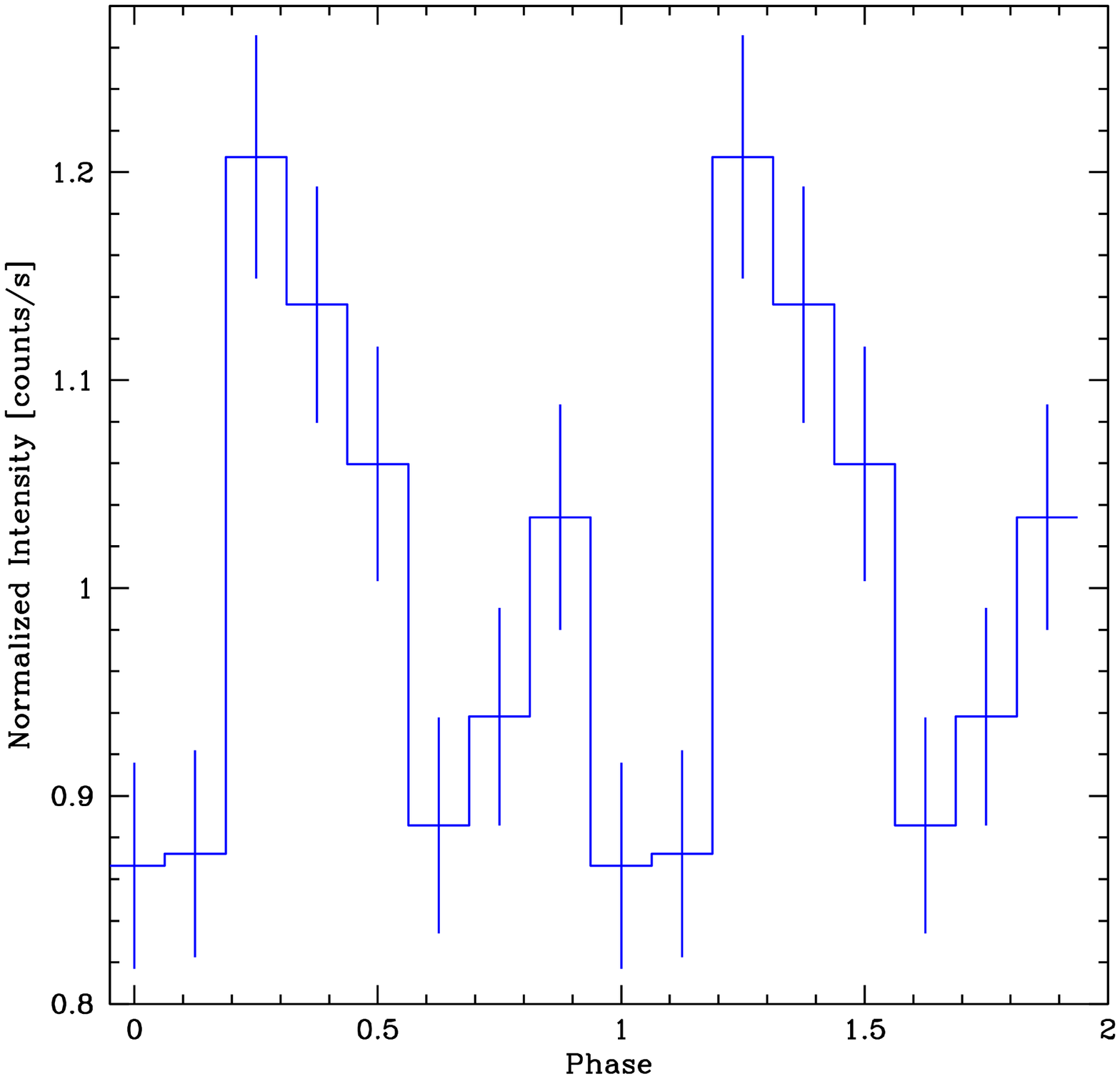}
\caption{\emph{Left}: Periodicity search ($\chi^{2}$ distribution) on the pn light curve (0.13--15\,keV) of IGR~J18532$+$0416 centered at 1408\,s (vertical line), with 8 bins per period, and a resolution of 5\,s. \emph{Right}: Pulse profile showing two phases of IGR~J18532$+$0416 for a period of 1408\,s. The zero phase (phase with the minimum flux) corresponds to MJD\,55663.00815.}
\label{lc_efsearch_18532}
\end{figure*}

\subsection{IGR~J18482$+$0049}

Our \xmm\ observation provides us with the most accurate X-ray position for \object{IGR~J18482$+$0049}: R.A. $= 18^{\mathrm{h}} 48^{\mathrm{m}} 15\overset{\mathrm{s}}{.}32$, and Dec. $= +00^{\circ} 47^{\prime} 34\overset{\prime\prime}{.}9$. This position is compatible with the discovery position of ISGRI \citep{bir10} which is $2\overset{\prime}{.}5$ away and has an error radius of $4\overset{\prime}{.}4$. \citet{ste10} proposed an association of \object{IGR~J18482$+$0049} with an object from the \xmm\ Slew Survey \citep[XMMSL:][]{sax08} that is located $2\overset{\prime}{.}2$ from the ISGRI position and has a 90\%-confidence radius of $13\overset{\prime\prime}{.}6$. However, our \xmm\ position is 21$^{\prime\prime}$ from the XMMSL source which is henceforth ruled out. Only one object in the 2MASS catalog is consistent with the \xmm\ position: \object{2MASS~J18481540$+$0047332}, which is located $2\overset{\prime\prime}{.}0$ from the center of the \xmm\ error circle and has magnitudes of $\ge$15.8, $\ge$14.0, and 13.9$\pm$0.1, respectively, in the $J$, $H$, and $K$ bands \citep{skr06}. This source is also listed as \object{GLIMPSE~G033.2747+01.0692} \citep{ben03}. There are no other objects from other wavelengths inside the \xmm\ error circle.

An absorbed power law is sufficient to model the X-ray spectrum of \object{IGR~J18482$+$0049} (Fig\,\ref{fig_spec}). The column density is large (\nh\ $\sim 4\times10^{23}$\,\cmsq) and is well in excess of the cumulative line-of-sight absorption value \citep{kal05}, and so favors an interpretation in which most of the obscuring material is intrinsic to the system. The photon index is $\Gamma =$ 2.0$\pm$0.3 with an unabsorbed flux of $7.6\times10^{-12}$\,\ergcms\ in the 0.5--10-keV band ($= 9.1\times10^{34} \left [ \frac{d}{10\textrm{ kpc}} \right ]^{2}$\,\ergs). This simple model provides an excellent fit ($\chi_{\nu}^{2}$/dof $=$ 1.01/108). An absorbed blackbody with a temperature of $1.9_{-0.1}^{+0.2}$\,keV also provides a good fit ($\chi_{\nu}^{2}$/dof $=$ 1.05/108), but residuals appear above 10\,keV. More complicated spectral models (i.e., cutoff power laws, and Comptonization) lead to unconstrained spectral parameters. We tested for the presence of a soft excess by adding a blackbody component to the absorbed power law. The addition of this component with a free temperature did not improve the fit, but holding the temperature at 0.1\,keV led to a marginal improvement over the absorbed power law ($\chi_{\nu}^{2}$/dof $=$ 0.99/107; F-test probability of 5\%) with the other spectral parameters (\nh\ and $\Gamma$) remaining consistent.

Given the shape of the spectral continuum and the large absorbing column, its location close to the Galactic Plane, and its persistent emission in the hard X-rays, we conclude that \object{IGR~J18482$+$0049} is most likely a new obscured HMXB in the Scutum Arm. Many sources in this class feature pulsation periods in the X-rays, but we did not find conclusive evidence for a periodic signal between 1\,s and 5000\,s in the the light curve of \object{IGR~J18482$+$0049}. Confirmation of the HMXB nature of \object{IGR~J18482$+$0049} must await spectral analysis of the candidate optical/IR counterpart proposed here.

\subsection{IGR~J18532$+$0416}

\citet{bir10} discovered \object{IGR~J18532$+$0416}, and the source was followed up by \swift-XRT which provided a refined X-ray position \citep{fio11}. The position that we obtain from our \xmm\ observation is R.A. $= 18^{\mathrm{h}} 53^{\mathrm{m}} 15\overset{\mathrm{s}}{.}83$, and Dec. $= +04^{\circ} 17^{\prime} 48\overset{\prime\prime}{.}5$. Figure\,\ref{fig_2mass} shows that this position is 7$\overset{\prime\prime}{.}$3 away from (but still statistically compatible with) the location of XRT Source \#2 in \citet{fio11}. The nearest potential counterpart across all wavelengths is \object{2MASS~J18531602$+$0417481} which is located $2\overset{\prime\prime}{.}9$ away. It has magnitudes of $\ge$16.5, 15.2$\pm$0.1, and 13.9$\pm$0.1 in the $J$, $H$, and $K$ bands, respectively \citep{skr06}.

A power law ($\Gamma =$ 1.4$\pm$0.1) with moderate absorption (\nh\ $=(1.98\pm0.08)\times10^{22}$\,\cmsq) provides a good fit to the source spectrum ($\chi_{\nu}^{2}$/dof $=$ 1.05/166). Residuals can be seen near 6\,keV (Fig.\,\ref{fig_spec}) so we included a Gaussian at $6.10_{-0.04}^{+0.06}$\,keV ($EW = 253_{-109}^{+122}$\,eV) which yields a fit with $\chi_{\nu}^{2}$/dof $=$ 0.94/163. This suggests a redshifted iron $K\alpha$ line source at $z = 0.05(1)$. The unabsorbed flux (0.5--10 keV) is $1.8\times10^{-12}$\,\ergcms\ which converts to a luminosity of $ 2.1\times10^{34} \left [ \frac{d}{10\textrm{ kpc}} \right ]^{2}$\,\ergs. The implied (unabsorbed) luminosity at $z = 0.05(1)$ would be $\sim10^{43}$\,\ergs\ ($H_{\circ} = 70$\,km\,s$^{-1}$\,Mpc$^{-1}$). Using a cutoff power law or a Compton thermalization model gave fits with acceptable $\chi_{\nu}^{2}$ values but left important parameters unconstrained (e.g., cutoff energy and plasma temperature). A blackbody model shows residuals at low and high energies and provides a poor fit overall. 

The light curve for \object{IGR~J18532$+$0416} is presented in Fig.\,\ref{fig_lc}. The source is soft with a hardness ratio that is negative during most of the observation. The $\chi^{2}$ distribution for the pn light curve (Fig.\,\ref{lc_efsearch_18532}) contains a weak potential period at 1408\,s ($\chi^{2} = 41$ for 8 bins per trial, i.e., 5-$\sigma$ significance, not corrected for the number of trials). This candidate signal is not present in the Lomb-Scargle periodogram nor in the power spectrum. If this periodic signal is real and coherent, it would suggest an accreting neutron star in an HMXB. On the other hand, if the iron line is real, then it would invalidate the pulsar hypothesis (ascribing the candidate signal to chance or to aperiodic variability on a 1-ks timescale) and would suggest either a redshifted line from material in a low-inclination disk around a black hole in an X-ray binary \citep[e.g.,][]{fab89,vdw89}, or a source located at a cosmological distance (i.e., an AGN). Just as with \object{IGR~J18457$+$0244}, if both the iron line and modulation are real, then the latter could be the signature of a low-frequency QPO from an AGN. In other words, we can not firmly establish the nature of \object{IGR~J18532$+$0416} because of the low signal-to-noise ratio of both the iron line and the periodicity. Confirmation (or refutation) of either of these observables will require additional observations.

\subsection{IGR~J18538$-$0102}

\object{IGR~J18538$-$0102} was listed as a new source in \citet{bir10}. \citet{ste10} noted the positional association with \object{G~32.1$-$0.9} \citep{fol97}, a candidate supernova remnant (SNR) located $0\overset{\prime}{.}6$ from the ISGRI position. Given that the spectrum of the IGR source is harder and more absorbed than that of the SNR, \citet{ste10} concluded that \object{IGR~J18538$-$0102} is probably a distant Galactic source or an extragalactic source, with only a coincidental association with \object{G~32.1$-$0.9}. Subsequently, an archival \xmm\ observation of the field which contained the source was reanalyzed by \citet{hal10} who found a persistent source at the following coordinates: R.A. $= 18^{\mathrm{h}} 53^{\mathrm{m}} 48\overset{\mathrm{s}}{.}50$, and Dec. $= -01^{\circ} 02^{\prime}  30\overset{\prime\prime}{.}0$. This source is also listed as \object{2XMM~J185348.4$-$010229} \citep{wat09}. Inside its error circle of $3\overset{\prime\prime}{.}2$, there is a single catalogued infrared source, \object{2MASS~J18534847$-$0102295}, which has magnitudes of $\ge$14.2, 14.0$\pm$0.1, and 12.5$\pm$0.1, respectively, in the $J$, $H$, and $K$ bands \citep{skr06}. Among multiple optical counterpart candidates present within the \xmm\ error circle, \citet{lut12} identified a faint source whose spectrum shows a broad H$\alpha$ line corresponding to a Sey-1 AGN at $z = 0.145(1)$.

Our X-ray position for \object{IGR~J18538$-$0102} is R.A. $= 18^{\mathrm{h}} 53^{\mathrm{m}} 48\overset{\mathrm{s}}{.}42$, and Dec. $= -01^{\circ} 02^{\prime}  28\overset{\prime\prime}{.}3$. This is $2\overset{\prime\prime}{.}1$ and $1\overset{\prime\prime}{.}5$ away from, but still compatible with, the \xmm\ positions of \citet{hal10} and \citet{mal10}, respectively. Compared to these previously-reported \xmm\ positions, our position is closer ($1\overset{\prime\prime}{.}2$ away) to the optical counterpart (Sey-1) identified by \citet{lut12}, and it remains compatible with the 2MASS source which is located $1\overset{\prime\prime}{.}6$ away, the optical/IR counterpart ($1\overset{\prime\prime}{.}4$ offset) listed in \citet{hal10}, and with \object{USNO-B1.0~0889$-$0406090} \citep{mon03}. With a $13\overset{\prime\prime}{.}6$ uncertainty radius \citep{sax08}, the position of the XMMSL source listed in \citet{ste10} encompasses our \xmm\ error circle (Fig.\,\ref{fig_2mass}). 

The X-ray spectrum of the source (Fig.\,\ref{fig_spec}) can be adequately described ($\chi_{\nu}^{2}$/dof $=$ 1.33/359) by a power law with moderate absorption: \nh\ $=(1.98\pm0.08)\times10^{22}$\,\cmsq. This is slightly larger than the \nh\ measured by \citet{hal10} and by \citet{mal10}, but the photon index ($\Gamma =$ 1.57$\pm$0.04) and unabsorbed 0.5--10-keV flux ($6.3\times10^{-12}$\,\ergcms) are in agreement. Adding a cutoff at $9_{-3}^{+5}$\,keV leads to a small improvement in the fit ($\chi_{\nu}^{2}$/dof $=$ 1.28/358). Replacing the power law with a radial blackbody or a thermal Comptonization model leads to large residuals and poor fits. 

No periodic signal could be found in the 1--5000-s range in the pn light curve of \object{IGR~J18538$-$0102}. The photon index that we measured is consistent with the AGN classification proposed by \citet{lut12}, and the column density is larger than the cumulative line-of-sight value indicating a distant or extragalactic source.

\section{Summary \& Conclusions}
\label{sec_sum}

In this work, we have analyzed \xmm\ observations of five unclassified hard X-ray sources located towards the Scutum Arm. Refined X-ray positions, soft X-ray light curves, and spectral energy distributions have been derived for all five targets (Tables\,\ref{tab_posn}--\ref{tab_spec}). 

Our results indicate that \igrtwo\ and \igrthree\ are probably new heavily-absorbed (\nh\ $> 10^{23}$\,\cmsq) high-mass X-ray binaries (HMXBs). The former is a slow X-ray pulsar ($P =$ 997$\pm$1\,s) in a supergiant fast X-ray transient (SFXT) system. This makes \igrtwo\ the seventh SFXT (in a class composed of $\sim$20 members) to have confirmed X-ray pulsations, providing a useful laboratory to test SFXT emission mechanisms which assume highly-magnetized ($B \gtrsim 10^{13}$\,G) neutron stars \citep[e.g.,][]{gre07b,boz08}. These two additional systems represent a 40\% increase in the number of known obscured HMXBs in this region (Fig.\,\ref{fig_gal_nh}), helping to reduce (somewhat) the asymmetry that we continue to observe between the large number of obscured HMXB systems found towards the Norma Arm compared with the low number found in the direction of the Scutum Arm \citep{bod07}. 

We can not conclusively determine the nature of the three other targets in our sample, \igrone, \igrfour, and \igrfive. We propose that \igrone\ and \igrfour\ are probably active galactic nuclei (AGN) viewed through the plane of the Milky Way, but Galactic X-ray binaries can not be ruled out. For \igrfive, the spectral parameters (photon index and column density) are consistent with an AGN interpretation.

Confirmation of the classifications for all sources in this study (see Table\,\ref{tab_class}) will require further observations of the optical/infrared counterparts which we propose here.

%__________________________________________________________________Types
%
\begin{table}[!t] 
\caption{Proposed classifications for the five sources in this study.}
\vspace{2mm}
\begin{tabular}{ l c c }
\hline
\hline
Source Name	      			& Classification		\\	
\hline
\object{IGR~J18457$+$0244}	& probable AGN, HMXB possible	 \\

\object{IGR~J18462$-$0223}	& absorbed SFXT pulsar	 \\	
	
\object{IGR~J18482$+$0049}	& absorbed HMXB	 \\		

\object{IGR~J18532$+$0416}	& AGN? HMXB?	 \\	
	
\object{IGR~J18538$-$0102}	& probable Sey-1	 \\		
\hline
\end{tabular}
\label{tab_class}
\end{table}

\acknowledgments
The authors thank the anonymous referee whose constructive review led to significant improvements in the manuscript. This research has made use of: observations obtained with XMM-Newton, an ESA science mission with instruments and contributions directly funded by ESA Member States and NASA; data obtained from the High Energy Astrophysics Science Archive Research Center (HEASARC) provided by NASA's Goddard Space Flight Center; the SIMBAD database operated at CDS, Strasbourg, France; NASA's Astrophysics Data System Bibliographic Services; the Two Micron All Sky Survey, which is a joint project of the University of Massachusetts and the Infrared Processing and Analysis Center/California Institute of Technology, funded by the National Aeronautics and Space Administration and the National Science Foundation; and the IGR Sources page maintained by JR and AB (\texttt{http://irfu.cea.fr/Sap/IGR-Sources}).

{\it Facilities:} \facility{XMM-Newton}.

\bibliographystyle{apj}
\bibliography{bodaghee.bib}
\clearpage

\end{document}